\documentclass[twocolumn]{aastex631}
\received{\today}
\shorttitle{LEGWORK}
\graphicspath{{./}{figures/}}

\usepackage{physics}
\usepackage{mathtools}
\usepackage{etoolbox}

\usepackage[hang,flushmargin]{footmisc} 

\usepackage{showyourwork}

\definecolor{WildStrawberry}{RGB}{239, 85, 159}
\newcommand{\referee}[1]{{#1}}

\makeatletter
\newcommand{\unit}[1]{%
    \,\mathrm{#1}\checknextarg}
\newcommand{\checknextarg}{\@ifnextchar\bgroup{\gobblenextarg}{}}
\newcommand{\gobblenextarg}[1]{\,\mathrm{#1}\@ifnextchar\bgroup{\gobblenextarg}{}}
\makeatother

\newcommand{\avg}[1]{\left\langle#1\right\rangle}

\definecolor{SeaGreen}{RGB}{63, 188, 157}
\colorlet{sywScriptUpToDate}{SeaGreen}

\newcommand{\lw}{LEGWORK}
\newcommand{\lwColour}{SeaGreen}

\newcommand{\lwModLink}[1]{\href{https://legwork.readthedocs.io/en/latest/modules.html\#module-legwork.#1}{\color{\lwColour}{\texttt{#1}}}}
\newcommand{\lwFuncURL}[2]{https://legwork.readthedocs.io/en/latest/api/legwork.#1.#2.html}

\newcommand{\docsIcon}{{\color{\lwColour}{\faFileCode}}}
\newcommand{\docsLink}[1]{\href{#1}{\docsIcon}}

\newcommand{\tutorialIcon}{{\color{\lwColour}{\faLaptopCode}}}
\newcommand{\tutorialLink}[1]{\href{#1}{\tutorialIcon}}

\newtagform{eqtag}[]{(}{)}

\renewenvironment{equation}[1]{%
    \ifstrempty{#1}{%
        \renewtagform{eqtag}[]{(}{)}%
    }{%
        \renewtagform{eqtag}[]{\docsLink{#1}\,(}{)}%
    }%
    \usetagform{eqtag}%
    \align%
    }{%
    \endalign%
    \renewtagform{eqtag}[]{(}{)}%
    \usetagform{eqtag}%
}

\begin{document}

\title{LEGWORK: A python package for computing the evolution and detectability \\ of stellar-origin gravitational-wave sources with space-based detectors}

\newcommand{\cfa}{Center for Astrophysics | Harvard \& Smithsonian, 60 Garden Street, Cambridge, MA 02138, USA}
\newcommand{\mpa}{Max-Planck-Institut für Astrophysik, Karl-Schwarzschild-Straße 1, 85741 Garching, Germany}
\newcommand{\cca}{Center for Computational Astrophysics, Flatiron Institute, 162 Fifth Ave, New York, NY, 10010, USA}
\newcommand{\UoA}{Anton Pannekoek Institute for Astronomy and GRAPPA, University of Amsterdam, NL-1090 GE Amsterdam, The Netherlands}
\newcommand{\UW}{Department of Astronomy, University of Washington, Seattle, WA, 98195}

\author[0000-0001-6147-5761]{T. Wagg}
\affiliation{\UW}
\affiliation{\cfa}
\affiliation{\mpa}

\author[0000-0001-5228-6598]{K. Breivik}
\affiliation{\cca}

\author[0000-0001-9336-2825]{S. E. de Mink}
\affiliation{\mpa}
\affiliation{\UoA}
\affiliation{\cfa}

\correspondingauthor{Tom Wagg}
\email{tomjwagg@gmail.com}

\begin{abstract}
    We present \lw{} (LISA Evolution and Gravitational Wave Orbit Kit), an open-source Python package for making predictions about stellar-origin gravitational wave sources and their detectability in LISA or other space-based gravitational wave detectors. \lw{} can be used to evolve the orbits of sources due to gravitational wave emission, calculate gravitational wave strains (using post-Newtonian approximations), compute signal-to-noise ratios and visualise the results. It can be applied to a variety of potential sources, including binaries consisting of white dwarfs, neutron stars and black holes. Although we focus on double compact objects, in principle \lw{} can be used for any system with a user-specified orbital evolution, such as those affected by a third object or gas drag. We optimised the package to make it efficient for use in population studies which can contain tens-of-millions of sources. This paper describes the package and presents several potential use cases. We explain in detail the derivations of the expressions behind the package as well as identify and clarify some discrepancies currently present in the literature. We hope that \lw{} will enable and accelerate future studies triggered by the rapidly growing interest in gravitational wave sources.\\

    \noindent Software reviewed by the \href{https://doi.org/10.21105/joss.03998}{Journal of Open Source Software} \citep{LEGWORK_joss}
\end{abstract}

\vspace{-1.25cm}

\keywords{open source software, gravitational waves, gravitational wave detectors, compact objects, orbital evolution, white dwarfs, neutron stars, stellar mass black holes }

\section{Introduction}

The planned space-based gravitational wave detector LISA (Laser Interferometer Space Antenna) will present an entirely new view of gravitational waves by focusing on lower frequencies ($10^{-5} < f / \unit{Hz} < 10^{-1}$) than ground-based detectors. This will enable the study of many new source classes including mergers of supermassive black holes \citep[e.g.][]{Begelman+1980, Klein+2016, Bellovary2019}, extreme mass ratio inspirals \citep[e.g.][]{Berti2006, Barack2007, Babak2017, Moore2017}, and cosmological GW backgrounds \citep[e.g.][]{Bartolo2016, Caprini2016, Caldwell2019}. However, this frequency regime is also of interest for the detection of local stellar-mass binaries during their inspiral phase. LISA is expected to detect Galactic stellar-origin binaries containing combinations of white dwarfs, neutron stars, and black holes, ranging from the numerous double white dwarf population, to the rare but loud double black hole population.

The potential to detect stellar-origin sources with LISA has been studied in the past by various groups \citep[e.g.][]{Nelemans+2001, Liu+2009, Liu+2014, Ruiter+2010, Belczynski+2010, Yu+2010, Nissanke+2012}. The recent direct detection of gravitational waves with ground-based detectors has led to renewed interest in this topic \citep[e.g.][]{Korol+2017, Korol+2018, Korol+2019, Korol+2020, Christian+2017, Kremer+2017, Kremer+2018, Lamberts+2018, Lamberts+2019, Fang+2019, Andrews+2020, Lau+2020, Breivik+2020, Breivik+2020a, Roebber+2020, Chen+2020, Sesana+2020, Shao+2021}.

Each of these studies require making estimates of the signal-to-noise ratio of individual binary systems and possibly the slow gravitational wave inspiral that lead to the present-day parameters. So far, most studies made use of custom made codes which have not been made publicly available. 

We believe that the large renewed interest in LISA and the stellar-origin sources it may detect will lead to many more studies in the near future that would need similar computations. This leads to a significant amount of redundancy which, at best results in extra work for each individual and at worst leads to an increased chance of introducing mistakes and inconsistencies when translating the necessary expressions to software.

\lw{} is an open-source Python package designed to streamline the process of making predictions of LISA detection rates for stellar-origin binaries such that it is as fast, reliable and simple as possible. With \lw{} one can evolve the orbits of a binary or a collection of binaries and calculate their strain amplitudes for any range of frequency harmonics. One can compute the sensitivity curve for LISA or other future gravitational wave detectors (e.g.\ TianQin's curve, or that of a custom instrument) and use it to compute the signal-to-noise ratio of a collection of sources. Furthermore, \lw{} provides tools to visualise all of the results with easy-to-use plotting functions. Finally, \lw{} is fully tested to check for consistency in the derivations described below.

Specifically, we implement the post-Newtonian expressions by \citet{Peters+1963} and \citet{Peters+1964} for the evolution of binary orbits due to the emission of gravitational waves, equations for the strain amplitudes and signal-to-noise ratios of binaries from various papers \citep[e.g.][]{Flanagan+1998, Finn+2000, Cornish2003, Barack+2004, Moore+2015} and approximations for the LISA and TianQin sensitivity curves given in \citet{Robson+2019} and \citet{Huang+2020} respectively. \referee{The post-Newtonian expressions are approximately of order 0.5 as they account for the orbital evolution from energy loss due to the emission of GWs, but don’t include other effects such as spin-orbit coupling. We find this is an excellent approximation for stellar-origin sources in the LISA band as any higher order terms are either zero or negligible.}

The open-source nature of the project means that new users as well as seasoned experts in the field can work together in a collaborative setting to consider new features and enhancements to the package as well as check the implementation. At the same time, with our thorough online documentation, derivations and tutorials, we hope \lw{} can make this functionality more accessible to the broader scientific community.

We note that \lw{} is not the only initiative of this kind. We highlight the ``Gravitational Wave Universe Toolbox'' presented by \citet{Yi+2022} which was developed to simulate observations on the GW universe with different detectors covering the full gravitational-wave spectrum and source classes. We also highlight ``GWPlotter'' by \citep{Moore+2015}, which provides an interactive plotting tool to compare the sensitivity of different gravitational wave detectors. \lw{} differs from the tools listed here as we have tried to provide tools that are optimised to rapidly make predictions for large populations of stellar-origin sources and we have focused on space-based detectors.

\lw{} has been developed with stellar-origin binary population studies in mind. We highlight two recent papers that use our package. \citet{Wagg+2021} investigates several populations of potential LISA sources (double black holes, black hole neutron stars and double neutron stars). They use \lw{} to evolve each synthesised source over the age of the Milky Way, make predictions about the LISA detectable population and explore how it varies with different binary physics assumptions. Additionally, \citet{Thiele+2021} examines the implications of assuming a metallicity-dependent binary fraction for the formation of close double white dwarfs (WDWDs) in the LISA frequency band. They use \lw{} to calculate the SNR of WDWDs and, in particular, apply a custom noise model that combines the LISA sensitivity curve from \citet{Robson+2019} with a new fit for the Galactic confusion noise based on their WDWD population.

\lw{} can be installed with pip or obtained from the GitHub repository \href{https://github.com/TeamLEGWORK/LEGWORK}{\color{\lwColour}\faGithub}\footnote{\url{https://github.com/TeamLEGWORK/LEGWORK}}. All examples shown in this paper and code to reproduce the figures are available in the repository. Instructions for installation and basic usage are provided in the online documentation\footnote{\url{https://legwork.readthedocs.io/en/latest/}} which contain the most up-to-date instructions.

The paper is organised as follows. In Section~\ref{sec:LEGWORK-overview} we give an overview of the capabilities of the \lw{} package and its various modules. We detail the derivations of the equations relevant for \lw{} in Section~\ref{sec:derivations}. In Section~\ref{sec:example-uses} we outline some example use cases of \lw{} to demonstrate its use. Finally, in Section~\ref{sec:summary} we conclude and summarise our work.

\section{Package Overview}\label{sec:LEGWORK-overview}

\begin{figure}
    \centering
    \includegraphics[width=\columnwidth]{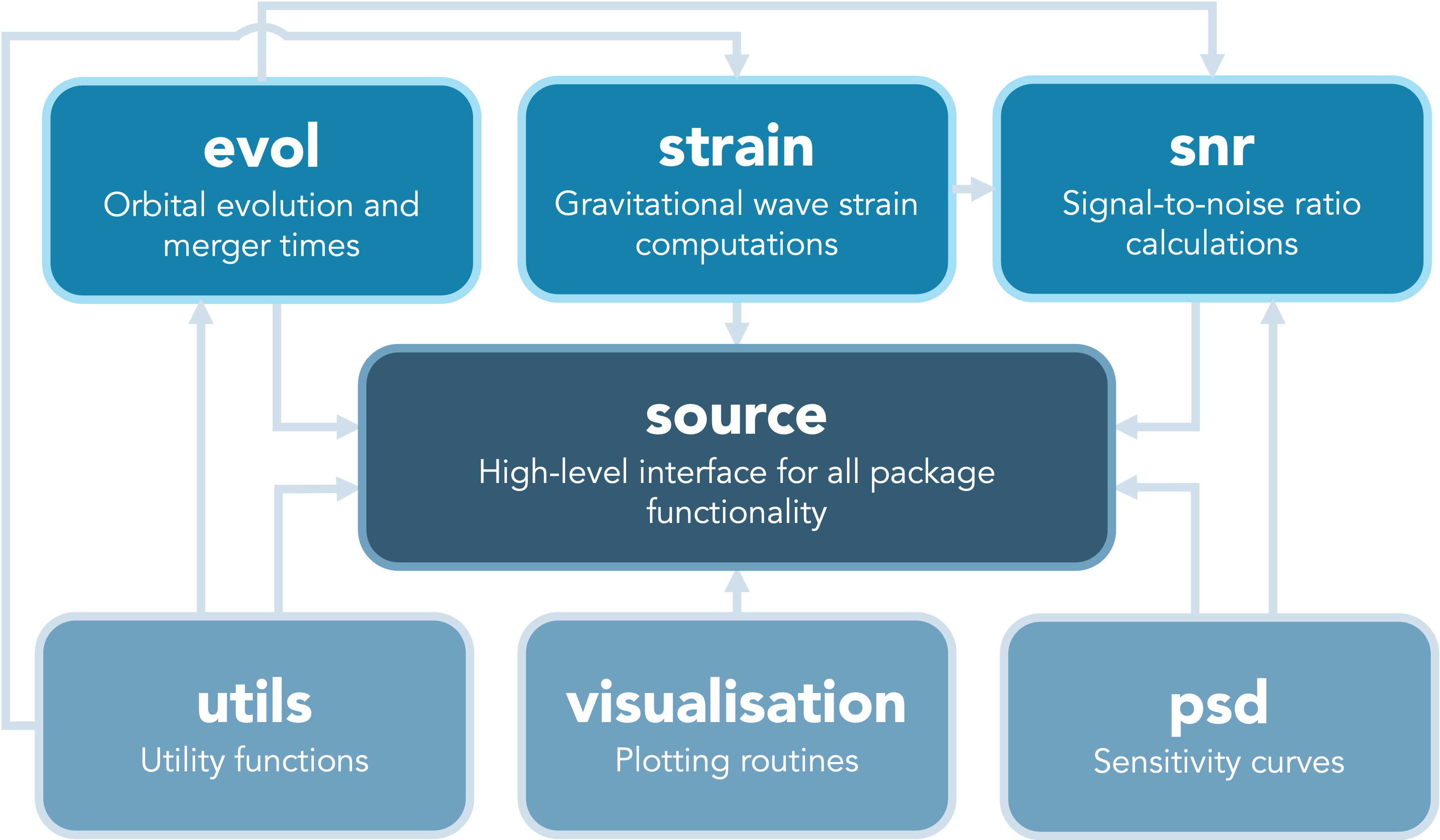}
    \caption{Package structure of \lw{}. Each box represents a module and describes its function. The arrows indicate the inter-dependencies of the modules.}
    \marginicon{\href{https://raw.githubusercontent.com/TeamLEGWORK/LEGWORK-paper/main/src/static/package_overview.png}{\color{\lwColour}\faFileImage}}
    \label{fig:package_overview}
\end{figure}

The \lw{} package is composed of seven modules that each focus on a particular aspect of calculations useful for gravitational wave sources that are detectable by space-based detectors. In Figure~\ref{fig:package_overview}, we illustrate the general structure of the package with each of its modules. The \texttt{source} module is the central module of the package and provides an simple interface to the functions in the rest of the modules. For more complex analyses, users may want to interact directly with individual modules, particularly those in the top row of Figure~\ref{fig:package_overview} as they comprise the core functionality of \lw{}. Below we explain the capabilities of each module in detail.

\lwModLink{evol} handles the orbital evolution of a binary due to the emission of gravitational waves. It includes functions for computing the merger times of both circular and eccentric binaries. In addition, you can use this module to evolve binary orbit parameters forward in time with any number or arrangement of timesteps. We discuss the relevant equations in Section~\ref{sec:deriv_evol}.

\lwModLink{strain} contains two functions that compute a binary's gravitational wave strain and characteristic strain amplitude respectively. Each of these functions is capable of computing the strain for an array of binaries at any number of timesteps and evaluated at any number of frequency harmonics. We discuss the relevant equations in Section~\ref{sec:deriv_strain}.

\lwModLink{psd} is used for evaluating the effective noise power spectral density of a detector at different frequencies. The module currently contains the LISA and TianQin sensitivity curves that can be tweaked by adjusting parameters such as the observation time, response function and even the arm length. \referee{For the Galactic confusion noise, a user can choose one of three models \citep{Robson+2019,Huang+2020,Thiele+2021}, use their own custom model, or turn it off entirely.} Additionally, this module allows the user to specify a custom detector sensitivity curve. We discuss the relevant equations in Section~\ref{sec:deriv_psd}.

\lwModLink{snr} uses the functions in \texttt{evol}, \texttt{strain} and \texttt{psd} to compute the signal-to-noise ratio of sources. It contains four functions that cover the permutations of whether a source is circular or eccentric and stationary in frequency space on the timescale of the mission or evolving. We discuss the relevant equations in Section~\ref{sec:deriv_snr}.

\lwModLink{visualisation} contains several wrappers for plotting 1- and 2-dimensional distributions with histograms, scatter plots, and kernel density estimator (KDE) plots in order to quickly analyse a collection of sources. In addition, it provides functions for plotting sources directly onto a sensitivity curve.

\lwModLink{source} provides a direct and simple interface to the functions in other modules through the \texttt{Source} Class. You can instantiate this Class with an array of sources and use it compute their strains or signal-to-noise ratios directly. Moreover, depending on the user's choice of allowed gravitational wave luminosity error, the Class dynamically decides on the number of frequency harmonics needed to capture the full signal of each binary and at what eccentricity to no longer consider a binary circular. This Class also provides a quick means of evolving the sources, visualising the parameters of each source and allows you to plot the binaries on the sensitivity curve.

\lwModLink{utils} is a collection of miscellaneous utility functions mainly consisting of conversions between variables as well as constants and expressions from \citet{Peters+1964}. We discuss the relevant equations in Section~\ref{sec:deriv_utils}.

\subsection{\referee{Units and automated testing}}

To ensure stability with physical units, all quantities included in \lw{} use the \texttt{astropy.units} module \citep{AstropyCollaboration+2013, AstropyCollaboration+2018}. This means that all inputs to \lw{} can be given in the units of the user's choice and will be automatically converted.

Furthermore, all of the source code in \lw{} is fully tested with continuous integration in the \lw{} GitHub repository. We employ several unit tests to ensure consistency between each of the use cases described below. For example, we require that the SNR calculation for circular and stationary binaries produces consistent output whether \lw{} uses the stationary and circular approximations or not. Similarly, we verify that the antenna patterns described below produce the expected values when averaged over source positions, inclinations, and polarisations.

\pagebreak

\subsection{Optimistations}

We developed \lw{} with an emphasis on increasing the efficiency of these computations in order to make simulations of large populations of systems tractable. We ensured that the entirety of \lw{} is vectorised and thus scales well with larger populations. In addition, we made a several specific optimisations to further increase the speed of calculations.

Firstly, we find that the runtime of calculating strains and SNRs for large populations of sources is mainly limited by the computation of (1) the relative gravitational-wave power in each harmonic for eccentric systems (see Eq.~\ref{eq:g(n,e)}) (2) the sensitivity curve of the given detector. Therefore, in order to significantly reduce the runtime of strain and SNR computations, by default \lw{} automatically interpolates these functions upon instantiation of any \texttt{Source} class with a large number of sources. Thereafter all functions use the tabulated values instead of calculating them exactly.

In certain cases one can apply approximations in place of the general SNR calculations. Although it is possible to use each of these approximations directly through the \lwModLink{snr} module, the \lwModLink{source} module will automatically apply the most appropriate function for each individual source. \lw{} dynamically classifies each source as one of four types, which cover the permutations of whether a binary is effectively circular or eccentric and whether or not it is stationary in frequency space on the timescale of the LISA mission. Then when computing the SNR, it applies a different function to each type of the source. This avoids computing unnecessary, time intensive integrals.

Moreover, for increasingly eccentric sources, gravitational waves are emitted in an increasing number of higher frequency harmonics. Although the total SNR is formally calculated as a the sum of the SNR over an infinite number of harmonics, in practice it is sufficient to only consider a subset. In the \lwModLink{source} module, the user can provide \texttt{gw\_lum\_tol}, a maximum allowable tolerance for the accuracy of the gravitational wave luminosity. Given this tolerance, \lw{} automatically calculates the required number of harmonics to satisfy this tolerance and thus minimise the computation time.

\section{Derivations}\label{sec:derivations}
In this section, we present a derivation of the equations used in \lw{}. We emphasise that these are not new derivations, conversely, they are in fact given frequently in the literature. However, they are often incomplete, unclear and, in some cases, contain spurious constant factors that arise from invalid combinations of previous work. Here, we aim to present a clear, clean and concise explanation of the expressions we use in \lw{}.

A \docsIcon{} symbol in an equation directly links to the relevant online \lw{} documentation for the implementation of that equation, which additionally contains a link to exact code used to reproduce the equation. These derivations are also given in more detail in the \lw{} documentation, where we show each of the intervening steps in more detail.

\subsection{Conversions and definitions {\normalfont (\texorpdfstring{\lwModLink{utils}}{utils})}}\label{sec:deriv_utils}
We start these derivations by defining some useful conversions and definitions. The chirp mass of a binary is the mass quantity measured by LISA and is given by
\begin{equation}{\lwFuncURL{utils}{chirp_mass}}
    \mathcal{M}_c \equiv \frac{(m_1 m_2)^{3/5}}{(m_1 + m_2)^{1/5}},
    \label{eq:chirpmass}
\end{equation}
where $m_1$ and $m_2$ are the primary and secondary masses of the binary. 

It is often convenient to convert between orbital frequency, $f_{\rm orb}$, and the semi-major axis, $a$, of a binary and this can be accomplished with Kepler's third law.
\begin{equation}{\lwFuncURL{utils}{get_a_from_f_orb}}
    a = \qty(\frac{G(m_1 + m_2)}{(2 \pi f_{\rm orb})^{2}})^{1/3},
    \label{eq:kepler3rd}
\end{equation}
where $G$ is the gravitational constant. Inversely,
\begin{equation}{\lwFuncURL{utils}{get_a_from_f_orb}}
    f_{\rm orb} = \frac{1}{2 \pi} \sqrt{\frac{G(m_1 + m_2)}{a^3}}.
    \label{eq:kepler3rd_reverse}
\end{equation}
For circular binaries, gravitational-wave emission occurs at twice the orbital frequency ($f_{\rm GW} = 2 f_{\rm orb}$). However, for eccentric binaries, we need to consider all frequency harmonics of gravitational-wave emission. These are defined such that the $n^{\rm th}$ harmonic frequency is
\begin{equation}{}
    f_{n} \equiv n \cdot f_{\rm orb}.
    \label{eq:nth_freq}
\end{equation}
It will be important to know the relative gravitational-wave power radiated into the $n^{\rm th}$ harmonic for a binary with eccentricity $e$ for the strain and SNR calculations. This is given by \citet[][Eq.\,20]{Peters+1963}
\begin{equation}{\lwFuncURL{utils}{peters_g}}
    \begin{aligned}
        g(n, e) = \frac{n^{4}}{32} & \Big\{ \Big[ J_{n-2}(n e)-2 e J_{n-1}(n e)+\frac{2}{n} J_{n}(n e) \\
        &\left. +2 e J_{n+1}(n e)-J_{n+2}(n e)\Big]^{2}\right.\\
        &\left.\kern-\nulldelimiterspace\right.
            \!\!\begin{aligned}
            +\big(1-e^{2}\big)\Big[&J_{n-2}(n e)-2 J_{n}(n e) \\
            &+J_{n+2}(n e)\Big]^{2}
            \end{aligned}\\
        &+\frac{4}{3 n^{2}}\Big[J_{n}(n e)\Big]^{2}\Big\},
    \end{aligned}
    \label{eq:g(n,e)}
\end{equation}
where $J_{n}(v)$ is the Bessel function of the first kind. Thus, the \textit{sum} of $g(n, e)$ over all harmonics gives the factor by which the gravitational-wave emission is stronger for a binary of eccentricity $e$ over an equivalent circular binary. This enhancement factor is \citep[][Eq.\,17]{Peters+1963}
\begin{equation}{\lwFuncURL{utils}{peters_f}}
    F(e) \equiv \sum_{n = 1}^{\infty} g(n, e) = \frac{1 + \frac{73}{24} e^2 + \frac{37}{96} e^4}{(1 - e^2)^{7/2}}.
    \label{eq:eccentricity_enhancement_factor}
\end{equation}
A useful rule of thumb is that $F(0.5) \approx 5.0$, or in words, a binary with eccentricity $0.5$ loses energy to gravitational waves at 5 times the rate of an equivalent circular binary.

\subsection{Orbital evolution {\normalfont (\texorpdfstring{\lwModLink{evol}}{evol})}}\label{sec:deriv_evol}
\subsubsection{Circular binaries}
For a circular binary, the orbital evolution due to gravitational-wave emission can be calculated analytically, as the rate at which the separation of the binary shrinks is simply a function of its mass and the current separation \citep[][Eq.~5.6]{Peters+1964}
\begin{equation}{}
    \dv{a}{t}_{e = 0} = -\frac{\beta}{a^3},
\end{equation}
where the constant $\beta$ is defined as
\begin{equation}{\lwFuncURL{utils}{beta}}
    \beta(m_1, m_2) \equiv \frac{64}{5} \frac{G^3}{c^5} m_1 m_2 (m_1 + m_2),
    \label{eq:beta_peters}
\end{equation}
where $c$ is the speed of light, $m_1$ is the primary mass and $m_2$ is the secondary mass. This gives the semi-major axis of a circular binary as a function of time, $t$, as \citep[][Eq.~5.9]{Peters+1964}
\begin{equation}{\lwFuncURL{evol}{evol_circ}}
    a(t, m_1, m_2) = [a_0^4 - 4 t \beta(m_1, m_2)]^{1/4},
    \label{eq:a_over_time_circ}
\end{equation}
where $a_0$ is the initial semi-major axis. Moreover, we can solve for the merger time, the time until the binary will merge, by setting the final semi-major axis in Eq.~\ref{eq:a_over_time_circ} \referee{to zero}.
\begin{equation}{\lwFuncURL{evol}{get_t_merge_circ}}
    t_{\rm merge, circ} = \frac{a_0^4}{4 \beta}.
    \label{eq:t_merge_circular}
\end{equation}

\subsubsection{Eccentric binaries}
The orbital evolution is more complex for eccentric binaries since the semi-major axis and eccentricity both evolve simultaneously and depend on one another. The final expression cannot be solved analytically and require numerical integration. The semi major axis, $a$, and eccentricity, $e$, are related as \citep[][Eq.~5.11]{Peters+1964}
\begin{equation}{\lwFuncURL{utils}{get_a_from_ecc}}
    a(e) = c_0 \frac{e^{12/19}}{(1 - e^2)} \left(1 + \frac{121}{304} e^2\right)^{870/2299},
    \label{eq:a_from_e}
\end{equation}
where $c_0$ satisfies the initial conditions such that $a(e_0) = a_0$. The time derivative of the eccentricity, $e$, is
\begin{equation}{\lwFuncURL{evol}{de_dt}}
    \frac{\mathrm{d}e}{\mathrm{d}t} = -\frac{19}{12} \frac{\beta}{c_{0}^{4}} \frac{e^{-29 / 19}\left(1-e^{2}\right)^{3 / 2}}{\left[1+(121 / 304) e^{2}\right]^{\frac{1181}{2299}}},
    \label{eq:dedt}
\end{equation}
which we can integrate to find $e(t)$ and convert to $a(t)$ using Eq.~\ref{eq:a_from_e} (\docsLink{\lwFuncURL{evol}{evol_ecc}}).

Inverting this function and applying the fact that we know that $e \to 0$ when the binary merges gives the merger time \citep[][Eq.~5.14]{Peters+1964}
\begin{equation}{\lwFuncURL{evol}{get_t_merge_ecc}}
    t_{\rm merge} = \frac{12}{19} \frac{c_{0}^{4}}{\beta} \int_0^{e_0} \frac{\left[1+(121 / 304) e^{2}\right]^{\frac{1181}{2299}}}{e^{-29 / 19}\left(1-e^{2}\right)^{3 / 2}} \mathrm{d}e.
    \label{eq:t_merge_eccentric}
\end{equation}
For very small or very large eccentricities we approximate this integral using the following expressions (given in unlabelled equations after \citealp[][Eq.~5.14]{Peters+1964})
\begin{align}
    t_{\rm merge,\, e^2 \ll 1} &= \frac{c_0^4}{4 \beta} \cdot e_0^{48 / 19}, \\
    t_{\rm merge,\, (1 - e^2) \ll 1} &= \frac{768}{425} \frac{a_0^4}{4 \beta} (1 - e_0^2)^{7/2}.
\end{align}
\noindent The standard threshold employed by \lw{} for small eccentricities is $e = 0.15$ and for large eccentricities is $e=0.9999$ (as this approximates $t_{\rm merge}$ with an error below roughly 2\%), though we note that this can be customised by the user if desired.

In addition, we implement the fit to Eq.~\ref{eq:t_merge_eccentric} from \citet{Mandel+2021} that approximates the merger time as
\begin{equation}{\lwFuncURL{evol}{t_merge_mandel_fit}}
    \begin{split}
        t_{\rm merge} \approx&\ t_{\rm merge, circ} (1 - e^2_0)^{7/2} \\
    &\times(1 + 0.27 e_0^{10} + 0.33 e^{20}_0 + 0.2 e_0^{1000} ),
    \end{split}
\end{equation}
which gives $t_{\rm merge}$ with an error below 3\% for eccentricities below $0.9999$. We additionally add a rudimentary polynomial fit to further reduce this error to below $0.5\%$. The user may specify whether to use this fit or perform the full integral when calculating merger times in \lw{}.

\subsection{Strains {\normalfont (\texorpdfstring{\lwModLink{strain}}{strain})}}\label{sec:deriv_strain}
\referee{\subsubsection{Characteristic Strain}
The strength of a gravitational wave in a detector at any one moment is determined by the strain amplitude, $h_0$. However, for stellar-origin sources at mHz frequencies, the signal can be present in the detector for many years. This means that, the $n^{\rm th}$ harmonic of the binary will spend approximately $f_n / \dot{f}_n$ seconds (or $f_n^2 / \dot{f}_n$ cycles) in the vicinity of a frequency $f_n$ \citep{Finn+2000}. This leads to the signal `accumulating' at the frequency $f_n$.

Therefore, to account for the integration of the signal over the mission, we instead use the `characteristic' strain amplitude of the $n^{\rm th}$ harmonic, $h_{c, n}$, which is the term present in the general signal-to-noise ratio equation. This can be related the to the strain amplitude in the $n^{\rm th}$ harmonic, $h_{0, n}$, as \citep[e.g][]{Finn+2000, Moore+2015}\footnote{Note that this is factor of 2 different from \citet{Finn+2000}. This is because the factor of 2 is already included in the \citet{Robson+2019} sensitivity curve and so is removed here.}
\begin{equation}{}
    h_{c, n}^2 = \qty(\frac{f_n^2}{\dot{f}_n}) h_{0, n}^2,
    \label{eq:strain-charstrain}
\end{equation}
The characteristic strain represents the strain measured by the detector over the duration of the mission (approximated as a single broad-band burst), whilst the strain amplitude is the strength of the GW emission at each instantaneous moment. For a stellar mass binary, the characteristic strain in the $n^{\rm th}$ harmonic is given by (e.g. \citealp[][Eq.\,56]{Barack+2004}; \citealp[][Eq.\,5.1]{ Flanagan+1998})}
\begin{equation}{}
    h_{c,n}^2 = \frac{1}{(\pi D_L)^2} \qty( \frac{2 G}{c^3} \frac{\dot{E_n}}{\dot{f_n}} ),
    \label{eq:char_strain_dedf}
\end{equation}
where $D_L$ is the luminosity distance to the source (note that for Milky Way sources, or any sources with redshift $\sim 0$, this is simply the distance to the source), $\dot{E}_n$ is the power radiated in the $n^{\rm th}$ harmonic and $\dot{f}_n$ is the rate of change of the $n^{\rm th}$ harmonic frequency.

The power radiated in the $n^{\rm th}$ harmonic can be expressed as \citep[][Eq. 19]{Peters+1963}
\begin{equation}{}
    \dot{E}_n = \frac{32}{5} \frac{G^{4}}{c^5} \frac{m_{1}^{2} m_{2}^{2}\left(m_{1}+m_{2}\right)}{a^{5}} g(n, e),
    \label{eq:edot_peters}
\end{equation}
where $g(n, e)$ is given in Eq.~\ref{eq:g(n,e)}. By substituting $a$ for $f_{\rm orb}$ (using Eq.~\ref{eq:kepler3rd_reverse}) and applying the definition of the chirp mass (Eq.~\ref{eq:chirpmass}) we obtain a more useful form for making gravitational wave predictions
\begin{equation}{}
    \dot{E}_n(\mathcal{M}_c, f_{\rm orb}, e) = \frac{32}{5} \frac{G^{7 / 3}}{c^{5}}\left(2 \pi f_{\mathrm{orb}} \mathcal{M}_{c}\right)^{10 / 3} g(n, e). \label{eq:edot}
\end{equation}
The last term needed to define the characteristic strain in Eq.~\ref{eq:char_strain_dedf} is the rate of change of the $n^{\rm th}$ harmonic frequency as a result of gravitational wave inspiral, which we can write as
\begin{equation}{}
    \dot{f}_{n} = \dv{f_{n}}{a} \dv{a}{t}.
    \label{eq:fdot_chainrule}
\end{equation}
We can find an expression for $\dd{f_{n}} / \dd{a}$ by substituting Eq.~\ref{eq:kepler3rd_reverse} into Eq.~\ref{eq:nth_freq} and differentiating 
\begin{equation}{}
    \dv{f_{n}}{a} = -\frac{3 n}{4 \pi} \frac{\sqrt{G(m_1 + m_2)}}{a^{5/2}}.
    \label{eq:dfda}
\end{equation}
The rate at which the semi-major axis decreases is \citep[][Eq. 5.6]{Peters+1964}
\begin{equation}{}
    \dv{a}{t} = -\frac{64}{5} \frac{G^{3} m_{1} m_{2}\left(m_{1}+m_{2}\right)}{c^{5} a^{3}} F(e).
    \label{eq:dadt}
\end{equation}
Substituting Eq.~\ref{eq:dfda} and Eq.~\ref{eq:dadt} into Eq.~\ref{eq:fdot_chainrule} gives an expression for $\dot{f}_{n}$
\begin{equation}{}
    \dot{f}_n = \frac{48 n}{5 \pi} \frac{G^{7/2}}{c^5} \qty(m_1 m_2 (m_1 + m_2)^{3/2}) \frac{F(e)}{a^{11/2}},
\end{equation}
which, as above with $\dot{E}_n$, we can recast using Kepler's third law and the definition of the chirp mass
\begin{equation}{\lwFuncURL{utils}{fn_dot}}
    \dot{f}_n = \frac{48 n}{5 \pi} \frac{\qty(G \mathcal{M}_c)^{5/3}}{c^5} (2 \pi f_{\rm orb})^{11/3} F(e).
    \label{eq:fdot}
\end{equation}
With definitions of both $\dot{E}_n$ and $\dot{f}_n$, we are now in a position to find an expression for the characteristic strain by plugging Eq.~\ref{eq:edot} and Eq.~\ref{eq:fdot} into Eq.~\ref{eq:char_strain_dedf}:
\begin{equation}{\lwFuncURL{strain}{h_c_n}}
    h_{c,n}^2 = \frac{2^{5/3}}{3 \pi^{4/3}} \frac{(G \mathcal{M}_c)^{5/3}}{c^3 D_L^2} \frac{1}{f_{\rm orb}^{1/3}} \frac{g(n, e)}{n F(e)}. \label{eq:char_strain}
\end{equation}

\subsubsection{Strain}
In order to obtain an expression for the strain amplitude of gravitational waves in the $n^{\rm th}$ harmonic, we can use Eq.~\ref{eq:strain-charstrain} and plug in Eq.~\ref{eq:fdot} and Eq.~\ref{eq:char_strain} 
\begin{equation}{\lwFuncURL{strain}{h_0_n}}
    h_n^2 = \frac{2^{28/3}}{5} \frac{(G \mathcal{M}_c)^{10/3}}{c^8 D_L^2} \frac{g(n, e)}{n^2} \qty(\pi f_{\rm orb})^{4/3} . 
    \label{eq:strain}
\end{equation}

\subsubsection{Amplitude modulation for orbit averaged sources}
\label{sec:amplitude_mod}
Because the LISA detectors are not stationary and instead follow an Earth-trailing orbit, the antenna pattern of LISA is not isotropically distributed or stationary. For sources that have unknown positions, inclinations, and polarisations, we use an average for the detector. However, for sources where these quantities are known we can consider the amplitude modulation of the strain due to the average motion of LISA's orbit.

We write that the position of the source on the sky is given by the ecliptic coordinates ($\theta$, $\phi$), the inclination of a source is $\iota$ and the polarisation of a source (determined by its orientation relative to the detector) is given by $\psi$. We follow the results of \citet{Cornish2003} to define the amplitude modulation. However, we adapt their expression to remain in the frequency domain and follow the conventions of more recent papers \citep[e.g.][Eq.\ 67]{Babak+2021} to write the amplitude modulation as

\begin{equation}{\lwFuncURL{strain}{amplitude_modulation}}
    A_{\rm{mod}}^{2}= \frac{1}{4} \left(1+\cos ^{2} \iota\right)^{2}\left\langle F_{+}^{2}\right\rangle_{\rm \referee{orb}}+ \cos ^{2} \iota\left\langle F_{\times}^{2}\right\rangle_{\rm \referee{orb}},
\label{eq:amp_mod}
\end{equation}
where $\left\langle F_{+}^{2}\right\rangle_{\rm \referee{orb}}$ and $\left\langle F_{\times}^{2}\right\rangle_{\rm \referee{orb}}$, the orbit-averaged detector responses, are defined as
\begin{equation}{\lwFuncURL{utils}{F_plus_squared}}
    \begin{split}
        \avg{ F_{+}^{2}}_{\rm \referee{orb}} = \frac{1}{4}\big(&\cos ^{2} 2 \psi\left\langle D_{+}^{2}\right\rangle_{\rm \referee{orb}} \\
        -&\sin 4 \psi\left\langle D_{+} D_{\times}\right\rangle_{\rm \referee{orb}}\\
        +&\sin ^{2} 2 \psi\left\langle D_{\times}^{2}\right\rangle_{\rm \referee{orb}}\big),
    \end{split}
\label{eq:response_fplus}
\end{equation}

\begin{equation}{\lwFuncURL{utils}{F_cross_squared}}
    \begin{split}
        \avg{ F_{\times}^{2}} = \frac{1}{4}\big(&\cos^{2} 2 \psi \avg{ D_{\times}^{2}}_{\rm \referee{orb}}\\
        +&\sin 4 \psi \avg{ D_{+} D_{\times}}_{\rm \referee{orb}}\\
        +&\sin ^{2} 2 \psi \avg{ D_{+}^{2}}_{\rm \referee{orb}} \big),
    \end{split}
\label{eq:response_fcross}
\end{equation}
and
\begin{equation}{\lwFuncURL{utils}{D_plus_D_cross}}
    \begin{split}
        \avg{ D_{+} D_{\times}}_{\rm \referee{orb}} = \frac{243}{512} \cos \theta \sin 2 \phi &\left(2 \cos ^{2} \phi-1\right) \\ &\times \left(1+\cos ^{2} \theta\right),
    \end{split}
\label{eq:d_plus_cross}
\end{equation}
\begin{equation}{\lwFuncURL{utils}{D_cross_squared}}
    \begin{split}
        \avg{ D_{\times}^{2}}_{\rm \referee{orb}} = \frac{3}{512}\big(120 \sin ^{2} \theta &+\cos ^{2} \theta \\
        &+162 \sin ^{2} 2 \phi \cos ^{2} \theta\big),
    \end{split}
\label{eq:d_cross}
\end{equation}
\begin{equation}{\lwFuncURL{utils}{D_plus_squared}}
    \begin{split}
        \avg{ D_{+}^{2} }_{\rm \referee{orb}} =  \frac{3}{2048}\big[&487+158 \cos ^{2} \theta+7 \cos ^{4} \theta\\
        &-162 \sin ^{2} 2 \phi\left(1+\cos ^{2} \theta\right)^{2}\big].
    \end{split}
\label{eq:d_plus}
\end{equation}
The orbital motion of LISA smears the source frequency by roughly $10^{-4}\,\rm{mHz}$ due to the antenna pattern changing as the detector orbits, the Doppler shift from the motion, and the phase modulation from the $+$ and $\cross$ polarisations in the antenna pattern. Generally, the modulation reduces the strain amplitude because the smearing in frequency reduces the amount of signal build up at the true source frequency.

\referee{We note that the amplitude modulation is only implemented in \lw{} for quasi-circular binaries to remain consistent with the calculation in \cite{Cornish2003}. Since the expected use case of LEGWORK is estimation of the detectability of large populations of mHz stellar-remnant binaries, for which predictions are uncertain by orders of magnitude in some cases, an extension of \cite{Cornish2003} which includes eccentric binaries is out of the current scope of \lw{}.}

\subsection{Sensitivity Curves {\normalfont (\texorpdfstring{\lwModLink{psd}}{psd})}}\label{sec:deriv_psd}

\subsubsection{LISA}
For the LISA sensitivity curve we use the equations from \citet{Robson+2019}. The \textit{effective} strain spectral density of the noise is defined as
\begin{equation}{\lwFuncURL{psd}{lisa_psd}}
    S_{\rm n}(f) \equiv \frac{P_n(f)}{\mathcal{R}(f)} + S_c(f),
\end{equation}
where $P_{\rm n}(f)$ is the power spectral density of the detector noise and $\mathcal{R}(f)$ is the sky and polarisation averaged signal response function of the instrument. Alternatively if we expand out $P_n(f)$, approximate $\mathcal{R}(f)$ and simplify we find \citep[][Eq.~1]{Robson+2019}
\begin{equation}{}
    \begin{split}
        S_{\rm n}(f) &= \frac{10}{3 L^2} \qty(P_{\rm OMS}(f) + 2\qty(1 + \cos^2 \qty(\frac{f}{f_*})) \frac{P_{\rm acc}(f)}{(2 \pi f)^4}) \\
        &\times \qty(1 + \frac{6}{10} \qty(\frac{f}{f_*})^2) + S_c(f),
    \end{split}
    \label{eq:LISA_Sn}
\end{equation}
where $L = 2.5 \unit{Gm}$ is detector arm length, $f_* = c / 2 \pi L = 19.09 \unit{mHz}$ is the transfer frequency, 
\begin{equation}{}
    \begin{split}
        P_{\rm OMS}(f) &= \left(1.5 \times 10^{-11} \mathrm{m}\right)^{2} \\
        &\times \left(1+\left(\frac{2 \mathrm{mHz}}{f}\right)^{4}\right) \mathrm{Hz}^{-1},
    \end{split}
\end{equation}
is the single-link optical metrology noise \citep[][Eq.~10]{Robson+2019},
\begin{equation}{}
    \begin{split}
        P_{\rm acc}(f) &= \left(3 \times 10^{-15} \mathrm{ms}^{-2}\right)^{2}\left(1+\left[\frac{0.4 \mathrm{mHz}}{f}\right]^{2}\right)\\
        &\times \left(1+\left[\frac{f}{8 \mathrm{mHz}}\right]^{4}\right) \mathrm{Hz}^{-1},
    \end{split}
\end{equation}
is the single test mass acceleration noise \citep[][Eq.~11]{Robson+2019} and
\begin{equation}{}
    \begin{split}
        S_{c}(f)&=A f^{-7 / 3} e^{-f^{\alpha}+\beta f \sin (\kappa f)} \\
        &\times \left[1+\tanh \left(\gamma\left(f_{k}-f\right)\right)\right] \mathrm{Hz}^{-1},
    \end{split}
\end{equation}
is the galactic confusion noise \citep[][Eq.~14]{Robson+2019}, where the amplitude $A$ is fixed as $9 \times 10^{-45}$ and the various parameters change over time and are listed in \citealt[][Table~1]{Robson+2019}. \lw{} allows the user to opt to use the \citet{Robson+2019} confusion noise, a custom function for the confusion noise or to remove the confusion noise entirely.

\subsubsection{TianQin}
For the TianQin sensitivity curve we use the power spectral density given in \citet[][Eq.~13]{Huang+2020}
\begin{equation}{\lwFuncURL{psd}{tianqin_psd}}
    \begin{split}
    S_{N}(f) &= \frac{10}{3 L^{2}}\left[\frac{4 S_{a}}{(2 \pi f)^{4}}\left(1+\frac{10^{-4} \unit{Hz}}{f}\right)+S_{x}\right] \\
    & \times\left[1+0.6\left(\frac{f}{f_{*}}\right)^{2}\right],
    \end{split}
\label{eq:tianqin}
\end{equation}
where $L = \sqrt{3} \times 10^5 \unit{km}$ is the arm length, $S_a = 1 \times 10^{-30} \unit{m^2}{s^{-4}}{Hz^{-1}}$ is the acceleration noise, $S_x = 1 \times 10^{-24} \unit{m^2}{Hz^{-1}}$ is the displacement measurement noise and $f_* = c / 2 \pi L$ is the transfer frequency. Note that Eq.~\ref{eq:tianqin} includes an extra factor of $10 / 3$ compared to \citet[][Eq.~13]{Huang+2020}. \citet{Huang+2020} absorb this factor into the waveform rather than include it in the power spectral density. We include it to match the convention used by \citet{Robson+2019} for the LISA sensitivity curve (see the factor of $10/3$ in Eq.~\ref{eq:LISA_Sn}) so that the sensitivity curves can be compared fairly.

\subsection{SNR for LISA {\normalfont (\texorpdfstring{\lwModLink{snr}}{snr})}}\label{sec:deriv_snr}
We note that this section draws heavily from \citet{Flanagan+1998} Section II C.

\subsubsection{Defining general SNR}
In order to calculate the signal to noise ratio for a given source of gravitational waves (GWs) in a 6-link LISA detector, we need to consider the following parameters:

\begin{itemize}
    \item position of the source on the sky: ($\theta$, $\phi$)
    \item direction from the source to the detector: ($\iota$, $\beta$)
    \item orientation of the source, which fixes the polarisation of the GW: $\psi$
    \item the distance from the source to the detector: $D_L$
\end{itemize}

Then, assuming a matched filter analysis of the GW signal $s(t) + n(t)$ (where $s(t)$ is the signal and $n(t)$ is the noise), which relies on knowing the shape of the signal, the signal to noise ratio, $\rho$, is given in the frequency domain as

\begin{align}{}
\label{eq:snr_general_start}
    \rho^2(D_L, \theta, \phi, \psi, \iota, \beta) &= \frac{\langle s(t)^{\star}s(t)\rangle}{\langle n(t)^{\star}n(t)\rangle}, \\
    &= 2 \int_{-\infty}^{+\infty} \frac{|\tilde{s}(f)|^2}{P_{\rm n}(f)} df, \\
    &= 4 \int_0^{\infty} \frac{|\tilde{s}(f)|^2}{P_{\rm n}(f)} df,
\end{align}
where $\tilde{s}(f)$ is the Fourier transform of the signal, $s(t)$, and $P_{\rm n}(f)$ is the one-sided power spectral density of the noise defined as as $\langle n(t)^{\star}n(t)\rangle = \int_0^{\infty} \frac{1}{2}P_{\rm n}(f) df$ \citep[c.f.][Eq.\,2]{Robson+2019}. Here, $\tilde{s}(f)$ is implicitly also dependent on $D_L, \theta, \phi, \psi, \iota,$ and $\beta$ as

\begin{equation}{}
\label{eq:signal}
    \begin{split}
        |\tilde{s}(f)|^2 = |&F_+(\theta, \phi, \psi)\tilde{h}_+(\referee{f}, D_L, \iota, \beta) \\
        + &F_{\times}(\theta, \phi, \psi)\tilde{h}_{\times}(\referee{f}, D_L, \iota, \beta)|^2,
    \end{split}
\end{equation}
where $F_{+,\times}$ are the `plus' and `cross' antenna patterns of the LISA detector to the `plus' and `cross' strains, $h_{+,\times}$. Note throughout any parameters discussed with the subscript $x_{+,\times}$ refers to both $x_{+}$ and $x_{\times}$.

\subsubsection{Average over position and polarisation}
\referee{Now, we can consider averaging over different quantities. In LISA's case, when averaged over all angles and polarisations, the antenna patterns are orthogonal and thus $\langle F_+ F_{\times}\rangle = 0$. This means we can rewrite Eq.~\ref{eq:signal} as 
\begin{equation}{}
    \begin{split}
        |\tilde{s}(f)|^2 &= |F_+(\theta, \phi, \psi)\tilde{h}_+(\referee{f}, D_L, \iota, \beta)|^2\\
        &+ |F_{\times}(\theta, \phi, \psi)\tilde{h}_{\times}(\referee{f}, D_L, \iota, \beta)|^2,
    \end{split}
\end{equation}
which can then be applied to Eq.~\ref{eq:snr_general_start} to give
\begin{equation}{}
\label{eq:position_orientation_ave}
    \langle \rho \rangle^2_{\theta,\phi,\psi} = 4 \int_0^{\infty} df \int \frac{d\Omega_{\theta,\phi}}{4\pi} \int \frac{d\psi}{\pi} \frac{|F_+\,\tilde{h}_+|^2 + |F_{\times}\,\tilde{h}_{\times}|^2}{P_{\rm n}(f)}.
\end{equation}}
From \cite{Robson+2019}, we can write the position and polarisation average of the signal response function of the instrument, $\mathcal{R}$, as

\begin{equation}{}
\label{eq:response}
    \begin{split}
        \mathcal{R} &= \langle F_+F^{\star}_+ \rangle = \langle F_{\times}F^{\star}_{\times} \rangle,\\
    \rm{where}\,\,
    \langle F_{+,\times}F^{\star}_{+,\times} \rangle &= \int \frac{d\Omega_{\theta,\phi}}{4 \pi} \int \frac{d\psi}{\pi} |F_{+,\times}|^2.
    \end{split}
\end{equation}
Then combining Eq.~\ref{eq:position_orientation_ave} and Eq.~\ref{eq:response}, we then find\footnote{Note that this is written in \cite{Flanagan+1998} for the LIGO response function which is $\mathcal{R} = \langle F_{+,\times} \rangle ^2 = 1/5$.}

\begin{equation}{}
\label{eq:averaged_antenna_simp}
    \langle \rho \rangle^2_{\theta,\phi,\psi} = 4 \int_0^{\infty} df \mathcal{R}(f)\, \left(\frac{|\tilde{h}_+|^2 + |\tilde{h}_{\times}|^2}{P_{\rm n}(f)}\right).
\end{equation}

\subsubsection{Average over orientation}
Now, we can average over the orientation of the source: $(\iota, \beta)$, noting that the averaging is independent of the distance $D_L$. With this in mind, we can rewrite $|\tilde{h}_+|^2 + |\tilde{h}_{\times}|^2$ in terms of two functions $|\tilde{H}_+|^2$  and  $|\tilde{H}_{\times}|^2$, where $\tilde{h}_{+,\times} = \tilde{H}_{+,\times}/D_L$. Given this, averaging over the source direction gives
\begin{equation}{}
\label{eq:averaged_all}
    \langle \rho \rangle^2_{(\theta,\phi,\psi),(\iota,\beta)} = \frac{4}{D_L^2} \int_0^{\infty} df \mathcal{R}(f)\,\int \frac{d\Omega_{\iota,\beta}}{4 \pi} \frac{|\tilde{H}_+|^2 + |\tilde{H}_{\times}|^2}{P_{\rm n}(f)},
\end{equation}
where we would like to express $\tilde{H}_{+,\times}(f)^2$ in terms of the energy spectrum of the GW. To do this, we note that the local energy flux of GWs at the detector is given by \citep[e.g.][Eq.\,6]{Press+1972}

\begin{equation}{}
\label{eq:energy_flx}
    \frac{dE}{dAdt} = \frac{1}{16\pi} \referee{\frac{c^3}{G}} \overline{\left[\left(\frac{dh_{+}}{dt}\right)^2 + \left(\frac{dh_{\times}}{dt}\right)^2\right]},
\end{equation}
where the bar indicates an average over several cycles of the wave which is appropriate for LISA sources. We can transform Eq.~\ref{eq:energy_flx} using Parseval's theorem, where we can write

\begin{align}
    \int_{-\infty}^{+\infty}dt\int dA \frac{dE}{dAdt} &= \int_{0}^{\infty}df \int dA \referee{\frac{c^3}{G}} \frac{\pi f^2}{2} \left(|\tilde{h}_{+}|^2 + |\tilde{h}_{\times}|^2 \right).
\label{eq:Parseval}
\end{align}
Note that the factor of frequency squared comes from the Fourier transform of the square of the time derivative in Eq.~\ref{eq:Parseval}. Now, since $A = D_L^2 \Omega$ and $|\tilde{h}_{+,\times}|^2 = |\tilde{H}_{+,\times}|^2 / D_L^2$, we find
\begin{equation}{}
\label{eq:little_h_to_big}
    |\tilde{h}_{+,\times}|^2 dA = |\tilde{H}_{+,\times}|^2 d\Omega_{\iota,\beta},
\end{equation}
then we can write Eq.~\ref{eq:Parseval} in terms of $|H_{+,\times}|^2$ as 
\begin{equation}{}
    \int_{-\infty}^{+\infty}dt\int dA \frac{dE}{dAdt} = \int_{0}^{\infty}df \frac{\pi f^2 \referee{c^3}}{2 \referee{G}} \int d\Omega \left(|\tilde{H}_{+}|^2 + |\tilde{H}_{\times}|^2 \right).
\label{eq:Parseval_2}
\end{equation}
Alternatively, by using Eq.~\ref{eq:little_h_to_big} and performing a Fourier transform we can also write that
\begin{equation}{}
    \int_{-\infty}^{+\infty}dt\int dA \frac{dE}{dAdt} = \int_{0}^{\infty}df \int d\Omega \frac{dE}{d\Omega df}.
    \label{eq:deriv_relation}
\end{equation}
From inspection of Eq.~\ref{eq:Parseval_2} and Eq.~\ref{eq:deriv_relation}, we can write the spectral energy flux as
\begin{equation}{}
    \int d\Omega \frac{dE}{d\Omega df} = \frac{\pi f^2 \referee{c^3}}{2 \referee{G}} \int d\Omega \left(|\tilde{H}_{+}|^2 + |\tilde{H}_{\times}|^2 \right) .
    \label{eq:se_flux}
\end{equation}

\subsubsection{Fulled averaged SNR}
We are now in a position to write an expression for the fully averaged SNR. Note for brevity we write $\langle \rho \rangle^2$ when referring to $\langle \rho \rangle^2_{(\theta,\phi,\psi),(\iota,\beta)}$. The application of Eq.~\ref{eq:se_flux} to Eq.~\ref{eq:averaged_all} yields
\begin{equation}{}
    \langle \rho \rangle^2 = \frac{4 \referee{G}}{\referee{c^3} D_L^2}\int_0^{\infty} df \frac{1}{P_{\rm n}(f)/\mathcal{R}(f)} \int \frac{d\Omega}{4\pi} \frac{dE}{d\Omega df} \frac{2}{\pi f^2}.
\end{equation}
This simplifies nicely to 
\begin{equation}{}
    \langle \rho \rangle^2 = \frac{2 \referee{G}}{\pi^2 \referee{c^3} D_L^2} \int_0^{\infty}df \frac{dE}{df}\frac{1}{f^2 P_{\rm n}(f)/\mathcal{R}(f)}.
\end{equation}
Finally, noting that $dE/df = dE/dt \times dt/df = \dot{E}/\dot{f}$, we can use the definition of the characteristic strain from Eq.~\ref{eq:char_strain_dedf} to finish up our position, direction, and orientation/polarisation averaged SNR as
\begin{equation}{}
    \langle \rho \rangle^2 = \int_0^{\infty}df \frac{h_{c}^2}{f^2P_{\rm n}(f)/\mathcal{R}(f)} = \int_0^{\infty}df \frac{h_{c}^2}{f^2 S_{\rm n}(f)},
    \label{eq:snr_finished_circ}
\end{equation}
where we have used that the effective power spectral density of the noise is defined as $S_{\rm n}(f) = P_{\rm n}(f) / \mathcal{R}(f)$.

It is also important to note that this is only the SNR for a circular binary for which we need only consider the $n = 2$ harmonic. In the general case, a binary could be eccentric and requires a sum over \textit{all} harmonics. Thus we can generalise Eq.~\ref{eq:snr_finished_circ} to eccentric binaries with
\begin{equation}{\lwFuncURL{snr}{snr_ecc_evolving}}
    \langle \rho \rangle^2 = \sum_{n = 1}^{\infty} \langle \rho_n \rangle^2 = \sum_{n = 1}^{\infty} \int_0^{\infty} d f_n \frac{h_{c, n}^2}{f_n^2 S_{\rm n}(f_n)},
    \label{eq:snr_general}
\end{equation}
where $h_{c, n}$ is defined in Eq.~\ref{eq:char_strain} and $S_{\rm n}$ in Eq.~\ref{eq:LISA_Sn}.

\subsubsection{SNR Approximations}
Although Eq.~\ref{eq:snr_general} can be used for every binary, it can be useful to consider different cases in which we can avoid unnecessary sums and integrals.
There are four possible cases for binaries in which we can use increasingly simple expressions for the signal-to-noise ratio. Binaries can be circular and stationary in frequency space. 

Circular binaries emit only in the $n=2$ harmonic and so the sum over harmonics can be removed. Stationary binaries have $f_{n, i} \approx f_{n, f}$ and so the small interval allows one to approximate the integral. Note we refer to non-stationary binaries as `evolving' here though many papers also use `chirping'.

For an evolving and eccentric binary, no approximation can be made and the SNR is found using Eq.~\ref{eq:snr_general}.

For a evolving and circular binary, the sum can be removed and so the SNR found as
\begin{equation}{\lwFuncURL{snr}{snr_circ_evolving}}
    \referee{\langle \rho \rangle^2_{\rm c, e}} = \int_{f_{2, i}}^{f_{2, f}} \frac{h_{c, 2}^2}{f_2^2 S_{\rm n}(f_2)} \dd{f_2}. \label{eq:snr_chirp_circ}
\end{equation}
For a stationary and eccentric binary we can approximate the integral.
\begin{align}
    \referee{\langle \rho \rangle^2_{\rm e, s}} &= \sum_{n=1}^{\infty}  \lim_{\Delta f \to 0} \int_{f_{n}}^{f_{n} + \Delta f_n} \frac{h_{c, n}^2}{f_n^2 S_{\rm n}(f_n)} \dd{f_n}, \\
    &= \sum_{n=1}^{\infty} \frac{\Delta f_n \cdot h_{c, n}^2}{f_n^2 S_{\rm n}(f_n)}, \\
    &= \sum_{n=1}^{\infty} \frac{\dot{f}_n \Delta T \cdot h_{c, n}^2}{f_n^2 S_{\rm n}(f_n)}, \\
    &= \sum_{n=1}^{\infty} \qty(\frac{\dot{f}_n}{f_n^2} h_{c, n}^2) \frac{T_{\rm obs}}{S_{\rm n}(f_n)},
\end{align}
where we have applied Eq.~\ref{eq:strain-charstrain} to convert between strains. This gives the following expression
\begin{equation}{\lwFuncURL{snr}{snr_ecc_stationary}}
    \referee{\langle \rho \rangle^2_{\rm e, s}} = \sum_{n=1}^{\infty} \frac{h_{n}^2 T_{\rm obs}}{S_{\rm n}(f_n)}. \label{eq:snr_stat_ecc}
\end{equation}
Finally, for a stationary and circular binary the signal-to-noise ratio is
\begin{equation}{\lwFuncURL{snr}{snr_circ_stationary}}
    \referee{\langle \rho \rangle^2_{\rm c, s}} = \frac{h_2^2 T_{\rm obs}}{S_{\rm n}(f_2)}.
    \label{eq:snr_stat_circ}
\end{equation}
For SNR calculations that take into account the amplitude modulation due to LISA's orbital motion (Section~\ref{sec:amplitude_mod}, we apply the calculations as described above but include the modulation in either the strain or characteristic strain as is appropriate.

\section{Use cases}\label{sec:example-uses}
In this section, we demonstrate \lw{}'s range of capabilities through a series of example use cases. The plots and results in each subsection are reproduced directly in online demos in the \lw{} documentation, which are each based on individual Jupyter notebooks. These tutorials are linked at the start of each subsection with the \tutorialIcon{} icon.

\subsection{Computing the SNR of a binary system\texorpdfstring{ - \tutorialLink{https://legwork.readthedocs.io/en/latest/demos/BasicSNRCalculation.html}}{}}

The most fundamental use case of \lw{} is to compute the SNR of an individual binary system. This can be accomplished in \lw{} with only two lines of code - one line to set up the source and another to compute the SNR. As an example one could consider a binary with the parameters
\begin{align*}
    m_1 = m_2 &= 10 \unit{M_{\odot}},\ d = 8 \unit{kpc},\\
    f_{\rm orb} &= 10^{-4} \unit{Hz},\ e = 0.2,
\end{align*}
where $m_1, m_2$ are the primary and secondary masses, $d$ is the distance to the source, $f_{\rm orb}$ is the orbital frequency and $e$ is the eccentricity. As we do not specify a position, polarisation or inclination, \lw{} will calculate the SNR averaged over these quantities. \referee{If a user specifies the position, then if either of the polarisation or inclination are not specified then they will be randomly generated. LEGWORK does not allow users to specify polarisations or inclinations without positions as we presume that the user must know the position if they know the polarisation or inclination of the source.} One can now instantiate a source in \lw{} using these parameters (leaving the mission parameters as the default values, such that we compute the SNR for a 4-year LISA mission). As shown in the linked demo, \lw{} quickly computes that the SNR of this binary is $4.49$. In the background, \lw{} decides which SNR approximation is most applicable given the eccentricity of the binary and whether it is stationary in frequency space.

This can be generalised to a population of many binary systems with ease. Instead of inputting single values for each parameter, one can input arrays of values where each entry corresponds to a different binary. As an example, we can take the same parameters as above for 3 different binaries but vary the primary as $m_1 = [5, 10, 15] \unit{M_{\odot}}$. Using \lw{} we find that the SNR for each of these cases is $\rho = [2.47, 4.49, 7.85]$. \referee{This also need not be limited to a 4-year LISA mission. With \lw{} we can additionally specify various parameters for the detector. For example, using \lw{} we can find that, for a 5-year TianQin mission with no confusion noise, the SNR for each of these cases is $\rho = [1.07, 1.95, 3.41]$.}

\subsection{Horizon distance\texorpdfstring{ - \tutorialLink{https://legwork.readthedocs.io/en/latest/demos/HorizonDistance.html}}{}}

\begin{figure*}[htb]
    \centering
    \includegraphics[width=\textwidth]{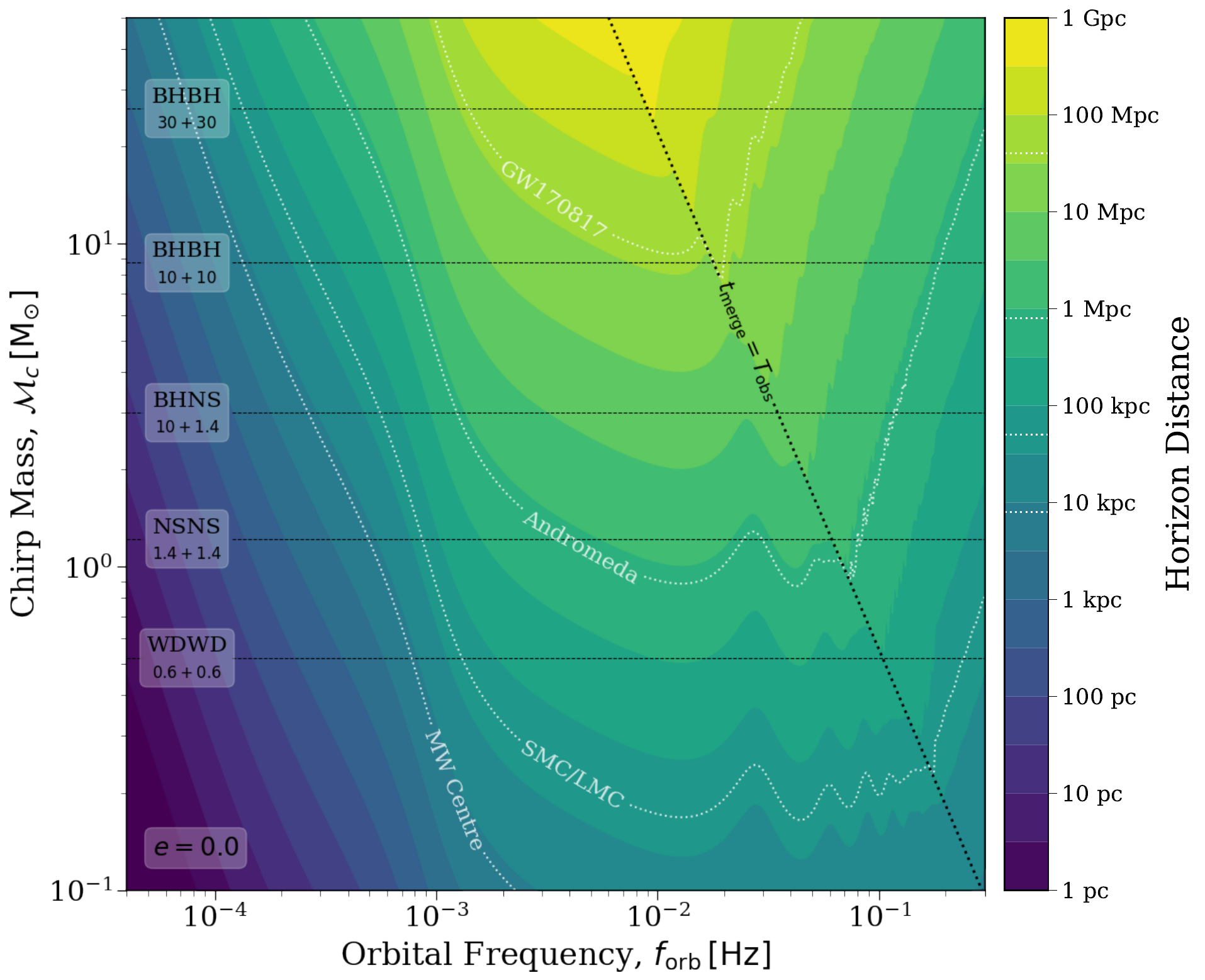}
    \caption{The horizon distance for circular, stellar-mass binaries in a 4-year LISA mission. The filled contours indicate the horizon distance for different orbital frequencies and chirp masses. We add white dotted contours at $8, 50, 800 \unit{kpc}$ and $40 \unit{Mpc}$ to highlight the distances to the centre of the Milky Way, the Magellanic Clouds, the Andromeda galaxy and the nearest ground-based gravitational wave detection (GW170817, \citealp{Abbott+2017_GW170817}) respectively. The diagonal black line shows the frequencies and chirp masses at which the merger time is equal to the observation time. This line emphasises the sharp contrast for the horizon distance for binaries that merge before the LISA mission finishes observing (to the right of the line). The horizontal black lines indicate the approximate location of some common double compact object types on this plot, with the assumed masses labelled below in solar masses.}
    \label{fig:horizon_distance}
    \script{horizon_distance.py}
\end{figure*}

A common question to consider with stellar mass sources in LISA is how far away a certain source could be detected. In other words, what is the horizon distance beyond which a source no longer has a SNR greater than some chosen threshold. We can explore this question using \lw{}.

Let us compute the horizon distance for a grid of chirp masses and orbital frequencies. First, we can recall that the signal-to-noise ratio of a source is inversely proportional to its distance from LISA and so we can find the horizon distance, $D_{\rm hor}$ as
\begin{equation}{}
    \label{eq:snr_to_hor_dist}
    D_{\rm hor} = \frac{\rho(D)}{\rho_{\rm detect}} \cdot D,
\end{equation}
where $\rho(D)$ is the signal-to-noise ratio as some distance $D$ and $\rho_{\rm detect}$ is the threshold above which we consider a source to be detectable. For the purpose of this example we will set $\rho_{\rm detect} = 7$. We can then, as a function of chirp mass and frequency, determine the maximum distance for which the source of interest is detected based on this SNR threshold.

This can be most efficiently accomplished using \lw{}'s \href{https://legwork.readthedocs.io/en/latest/api/legwork.source.Source}{\color{\lwColour}{\texttt{Source}}} class, since \lw{} can then compute the merger times and signal-to-noise ratio of each source with only two lines of code. We convert the SNRs to horizon distances using Eq.~\ref{eq:snr_to_hor_dist} and plot the result in Figure~\ref{fig:horizon_distance}. The shape of the LISA sensitivity curve is clearly reflected in the horizon distance. This is because circular sources only emit at a single frequency and thus every feature in the sensitivity curve has a strong effect on the SNR (and therefore horizon distance) for circular sources. We also see that once the merger time of a source is shorter than the LISA mission length (shown by the black dotted line), the horizon distance sharply decreases. This is because if a source merges before the LISA mission concludes, it has less time to accumulate signal and thus has a lower SNR and horizon distance.

One can use Figure~\ref{fig:horizon_distance} to estimate the horizon distance for any circular source of interest. To illustrate this, we add solid black lines to indicate the typical chirp masses of some possible stellar mass gravitational wave sources as well as white dotted lines to show the distances to nearby galaxies as well as the nearest ground-based gravitational wave detection (GW170817, \citealp{Abbott+2017_GW170817}). For example, we see that a circular NSNS with an orbital frequency greater than a mHz is detectable in the Magellanic clouds.

\subsection{The role of eccentricity\texorpdfstring{ - \tutorialLink{https://legwork.readthedocs.io/en/latest/demos/TheRoleofEccentricity.html}}{}}\label{sec:eccentricity_role}

The role of eccentricity is important to consider in the detection of gravitational waves with LISA and other space-based detectors, as sources can still have significant eccentricity during their inspiral phase. A high eccentricity has two major effects on the SNR of a gravitational wave source in LISA and we can investigate these effects using \lw{}.

Let's consider three hypothetical systems that are identical apart from their eccentricities, $e_i$.
\begin{align*}{}
    m_1 &= m_2 = 0.6 \unit{M_{\odot}}, f_{\rm orb} = 1.5 \unit{mHz}, d = 15 \unit{kpc}, \\
    e_i &= \{0.0, 0.6, 0.9 \},
\end{align*}
Using \lw{} to calculate their signal-to-noise ratios in a 4-year LISA mission, $\rho_i$, we find
\begin{equation*}{}
    \rho_i = \{ 31.7, 50.2, 38.8 \}.
\end{equation*}

We see two effects in the signal-to-noise ratio here. First, increasing the eccentricity from essentially circular to $e = 0.6$ results in a higher signal-to-noise ratio ($\rho=31.7 \to \rho=50.2$). This is because an eccentric binary has enhanced energy emission via gravitational waves \citep{Peters+1963}. This means that an eccentric binary will not only inspiral faster than an otherwise identical circular binary, but also will always produce a stronger gravitational wave strain. We discuss the enhancement factor and its exact dependence on eccentricity in more detail in Section~\ref{sec:derivations} (see specifically Eq.~\ref{eq:eccentricity_enhancement_factor}).

The second effect is more intriguing. We see that increasing the eccentricity from $e = 0.6$ to $e = 0.9$ results in a relative \textit{decrease} in SNR ($\rho=50.2 \to \rho=38.8$). The reason for this is that eccentric binaries emit gravitational waves at many harmonic frequencies (unlike circular binaries, which emit predominantly twice the orbital frequency). This leads to the gravitational wave signal being diluted over many frequencies higher than the orbital frequency, where the higher the eccentricity, the more harmonics are required to capture all of the gravitational luminosity \citep[see Figure~3 of][]{Peters+1963}. Therefore, if the eccentricity is too high, the majority of the signal may be emitted at a frequency to which LISA is less sensitive.

\begin{figure}[tb]
    \centering
    \includegraphics[width=\columnwidth]{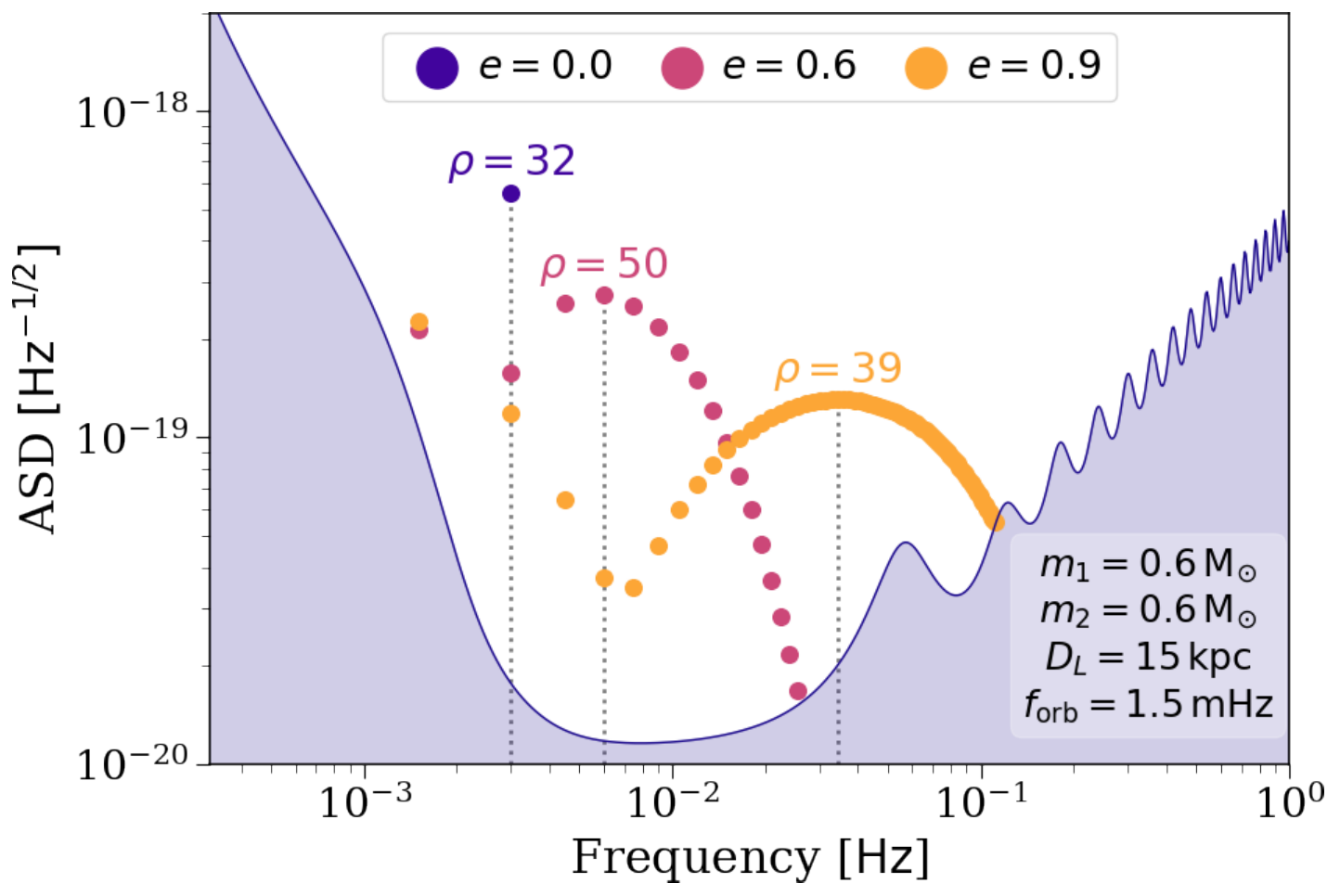}
    \caption{An illustration of the effect of eccentricity on the detectability of a LISA source. The three sets of points are coloured by their eccentricity and each individual point corresponds to a harmonic frequency, where its height above the curve gives its SNR. We annotate each set of points with its total SNR and overlay the LISA sensitivity curve. The dotted vertical lines indicate the frequency at which the majority of the gravitational wave signal is concentrated.}
    \label{fig:role_eccentricity}
    \script{role_eccentricity.py}
\end{figure}

We can illustrate this point with \lw{} by calculating the SNR at each individual frequency harmonic using the \lwModLink{snr} module. We plot this distribution of signal over different frequency harmonics with the LISA sensitivity curve overlaid in Figure~\ref{fig:role_eccentricity} using \lw{}'s \lwModLink{visualisation} module. A point is plotted for each harmonic of each source that has an SNR greater than unity and such that its height above the sensitivity curve corresponds to its SNR.

From Figure~\ref{fig:role_eccentricity}, we can better understand why a source with $e = 0.9$ has a lower SNR than the same source with $e = 0.6$. From the dotted lines, we can note that the signal from the $e = 0.9$ source is concentrated at a frequency of around $40 \unit{mHz}$. The LISA sensitivity at this point is much weaker than the $6 \unit{mHz}$ at which the $e =0.6$ source is concentrated. Therefore, although the strain from a more eccentric binary is stronger, the SNR is lower due to the increased noise in the LISA detector.

Overall, we can therefore conclude that for LISA sources of this nature, higher eccentricity will produce more detectable binaries only if the orbital frequency is not already at or above the minimum of the LISA sensitivity curve.

Another consideration for more massive binaries is whether the increased eccentricity will cause the binary to merge before the mission ends, which would cause a significant decrease in signal-to-noise ratio. We can also use \lw{} to find how the merger time of a source varies with frequency and eccentricity over a grid of sources.

\begin{figure}[htb]
    \centering
    \includegraphics[width=\columnwidth]{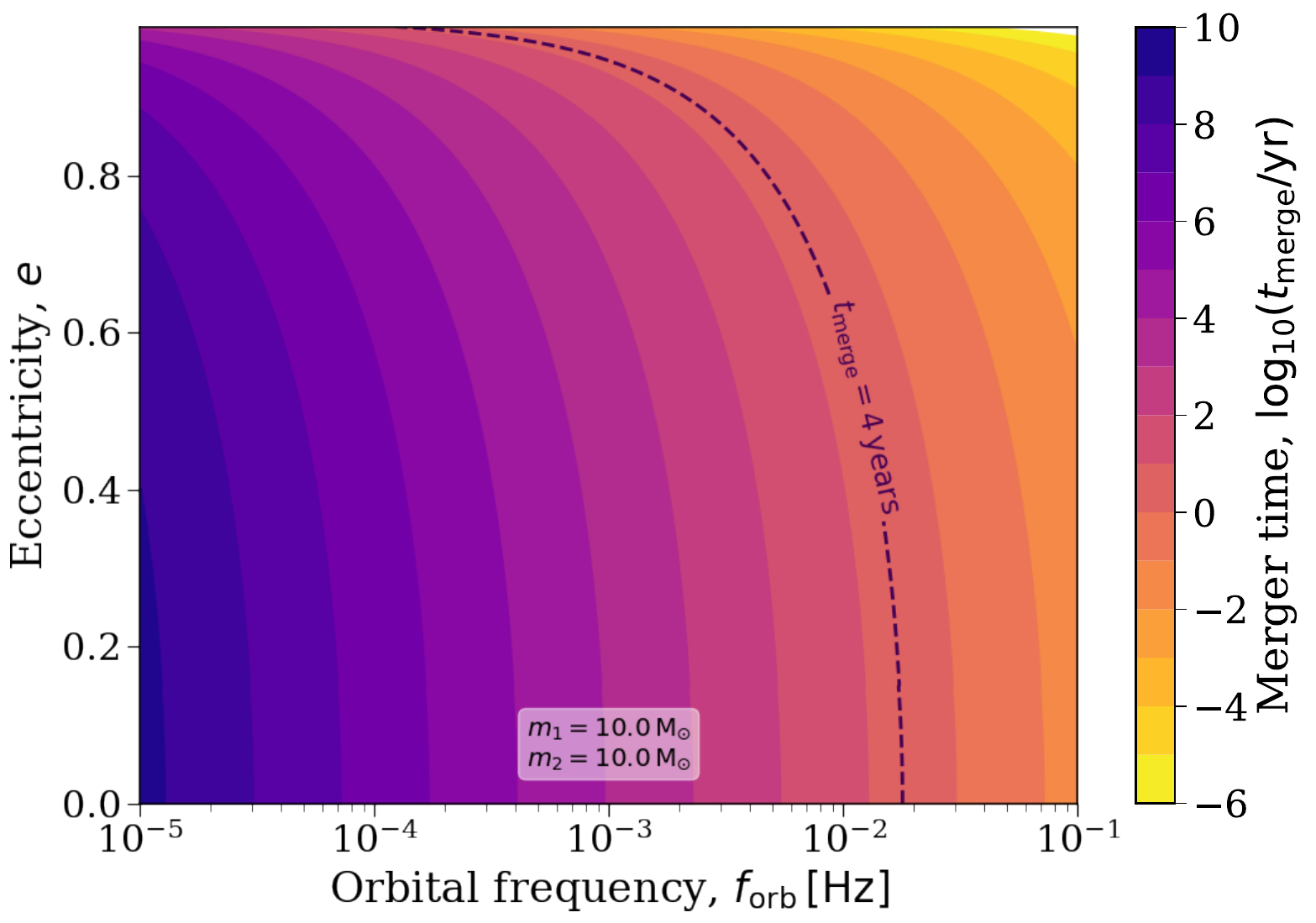}
    \caption{The merger time for a binary with both a primary and secondary mass of $10 \unit{M_{\odot}}$ over different eccentricities and orbital frequencies. The dashed line indicates a merger time of 4 years, the default LISA mission length.}
    \label{fig:merger_time}
    \script{merger_time.py}
\end{figure}

We plot the results of this calculation in Figure~\ref{fig:merger_time}. This plot shows that, for most eccentricities, the merger time is largely determined by the orbital frequency. However, for high eccentricities ($e > 0.8$), the eccentricity leads to a significant reduction in the merger time. Additionally, we can see that any binary that is to the right of the dashed line on this plot at the start of a 4-year LISA mission, would merge before the mission ended. Therefore, if increasing a binary's eccentricity moved it to the right of this line, its SNR will decrease significantly.

\subsection{Comparing gravitational wave detectors\texorpdfstring{ - \tutorialLink{https://legwork.readthedocs.io/en/latest/demos/CompareSensitivityCurves.html}}{}}

It may also be useful to consider how changing the specifications of the LISA detector, or using a different detector entirely, could affect the SNR of a particular source. \lw{} is capable of adjusting the LISA mission specifications or using a different sensitivity curve and thus we can use it to explore these differences.

\begin{figure}[tb]
    \centering
    \includegraphics[width=\columnwidth]{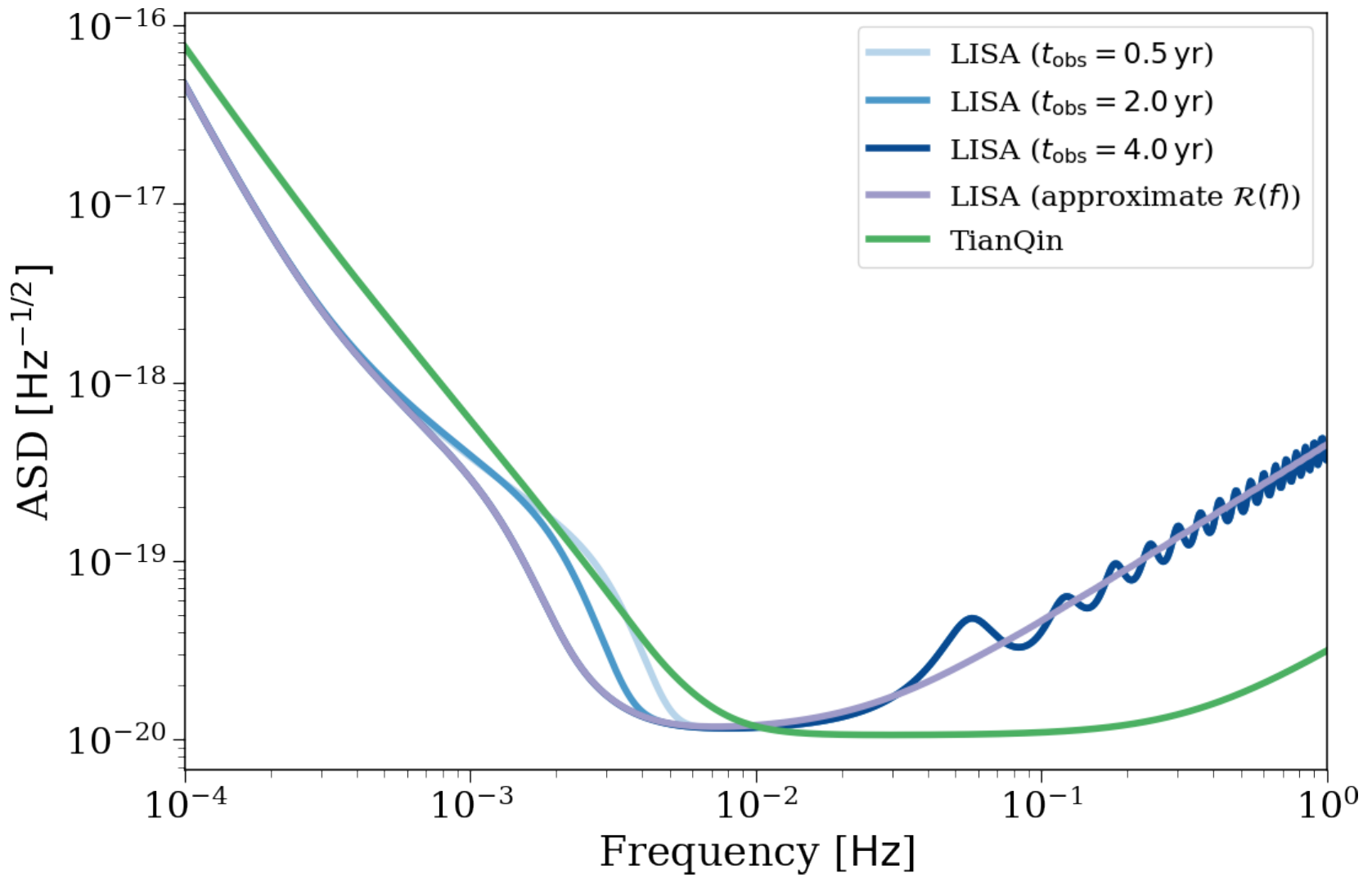}
    \caption{The strain spectral density of the LISA detector with different specifications \citep{Robson+2019} and the TianQin detector \citep{Huang+2020}. We show the LISA curve for three different mission lengths and once with an approximate response function.}
    \label{fig:detector_sc_compare}
    \script{detector_sc_compare.py}
\end{figure}

As a first step, we can use \lw{} to plot a series of sensitivity curves in Figure~\ref{fig:detector_sc_compare}. We show the LISA sensitivity curve for the default 4 year mission length but also illustrate how the curve changes for shorter mission lengths. At 0.5 and 2 years, we see a stronger noise level around $3 \unit{mHz}$ as a result of the increased Galactic confusion noise. This noise decreases with increasing mission length since more individual foreground sources can be resolved and thus removed from the confusion noise. We also see that using an approximated response function smooths out the sensitivity curve at higher frequencies. Finally, the TianQin curve is higher than the 4 year LISA curve until around $5 \unit{mHz}$, beyond which it has a lower noise level.

Although comparing the sensitivity curves would suffice for a stationary and circular source (since it would remain at a single frequency), \lw{} can also be used to see how the relative SNR between two detectors changes over a range of eccentricities and frequencies.

\begin{figure}[htb]
    \centering
    \includegraphics[width=\columnwidth]{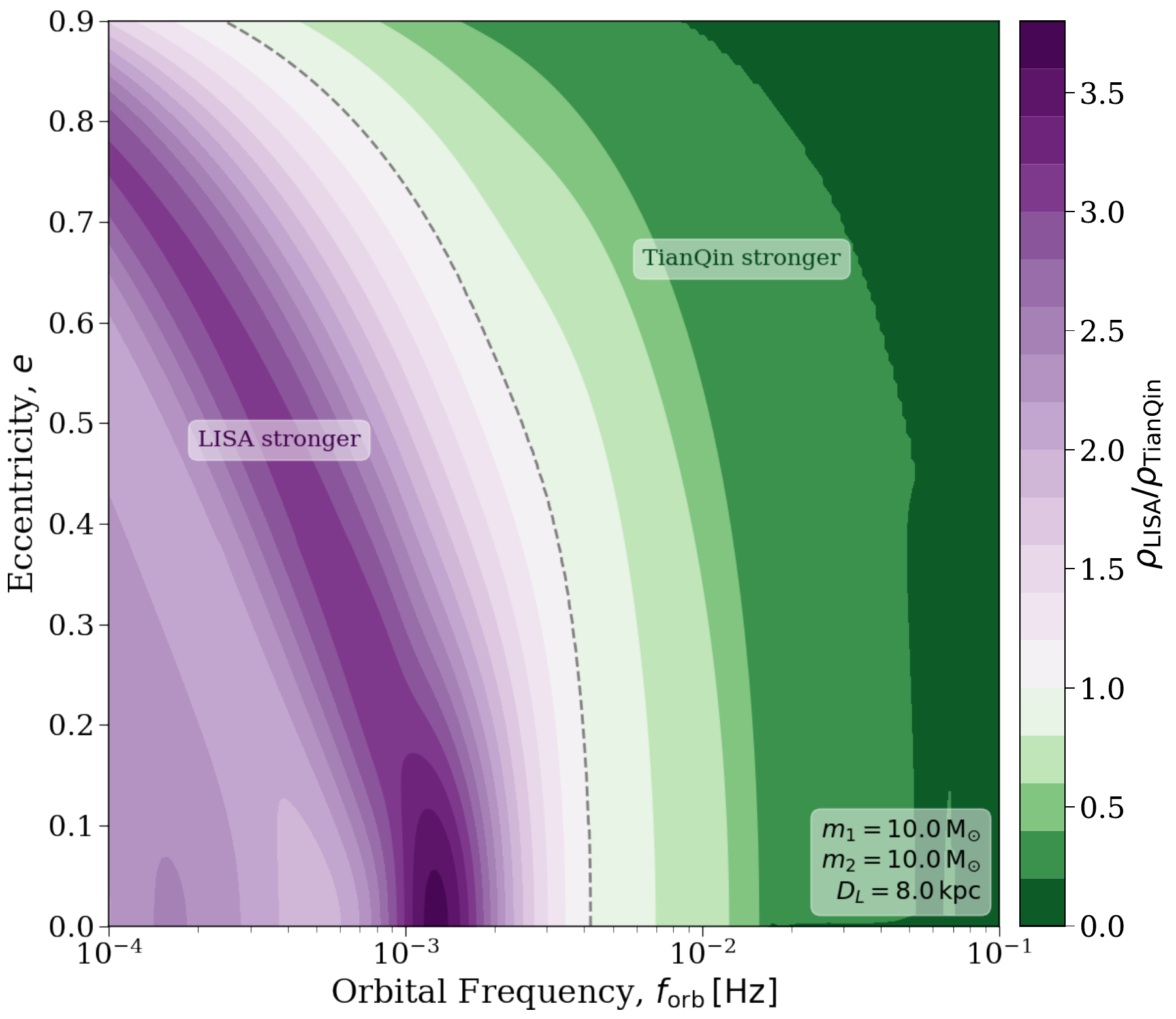}
    \caption{The ratio of the SNR in LISA (for a 4-year mission) to the SNR in TianQin. The dashed line indicates the transition at which the SNR is equal in both detectors. We annotate the regions in which either detector has a higher SNR and also annotate the mass and distance of each source in the grid.}
    \label{fig:detector_snr_ratio}
    \script{detector_snr_ratio.py}
\end{figure}

Using \lw{}, we can compute the SNR of a grid of sources (spanning a range of frequencies and eccentricities) for both detectors. In Figure~\ref{fig:detector_snr_ratio}, we show the ratio of the SNR in LISA to the SNR in TianQin. This plot shows that for circular binaries, the SNR of the source in LISA is stronger up to an orbital frequency of approximately $2.5 \unit{mHz}$, beyond which the SNR of the source is stronger in TianQin. This transition frequency becomes lower with increasing eccentricity as one would expect since eccentric sources emit more at higher harmonics and thus higher frequencies.

\subsection{Track SNR of a binary over time\texorpdfstring{ - \tutorialLink{https://legwork.readthedocs.io/en/latest/demos/SNROverTime.html}}{}}

As a binary inspirals its orbital frequency and eccentricity change and this in turn affects the SNR of the binary. For this use case we will demonstrate how \lw{} can be used to track the evolution of these parameters and pinpoint the moment at which a binary becomes detectable.

Let's consider a binary with the following initial parameters
\begin{align*}
    m_1 &= m_2 = 15 \unit{M_{\odot}},\ d = 20 \unit{kpc},\\
    f_{\rm orb, i} &= 3 \times 10^{-5} \unit{Hz},\ e_i = 0.5,
\end{align*}
and use \lw{}'s \lwModLink{evol} module to evolve the system until 100 years before its merger with 1000 linearly spaced timesteps, recording the eccentricity and frequency at each timestep.

\begin{figure}[htb]
    \centering
    \includegraphics[width=\columnwidth]{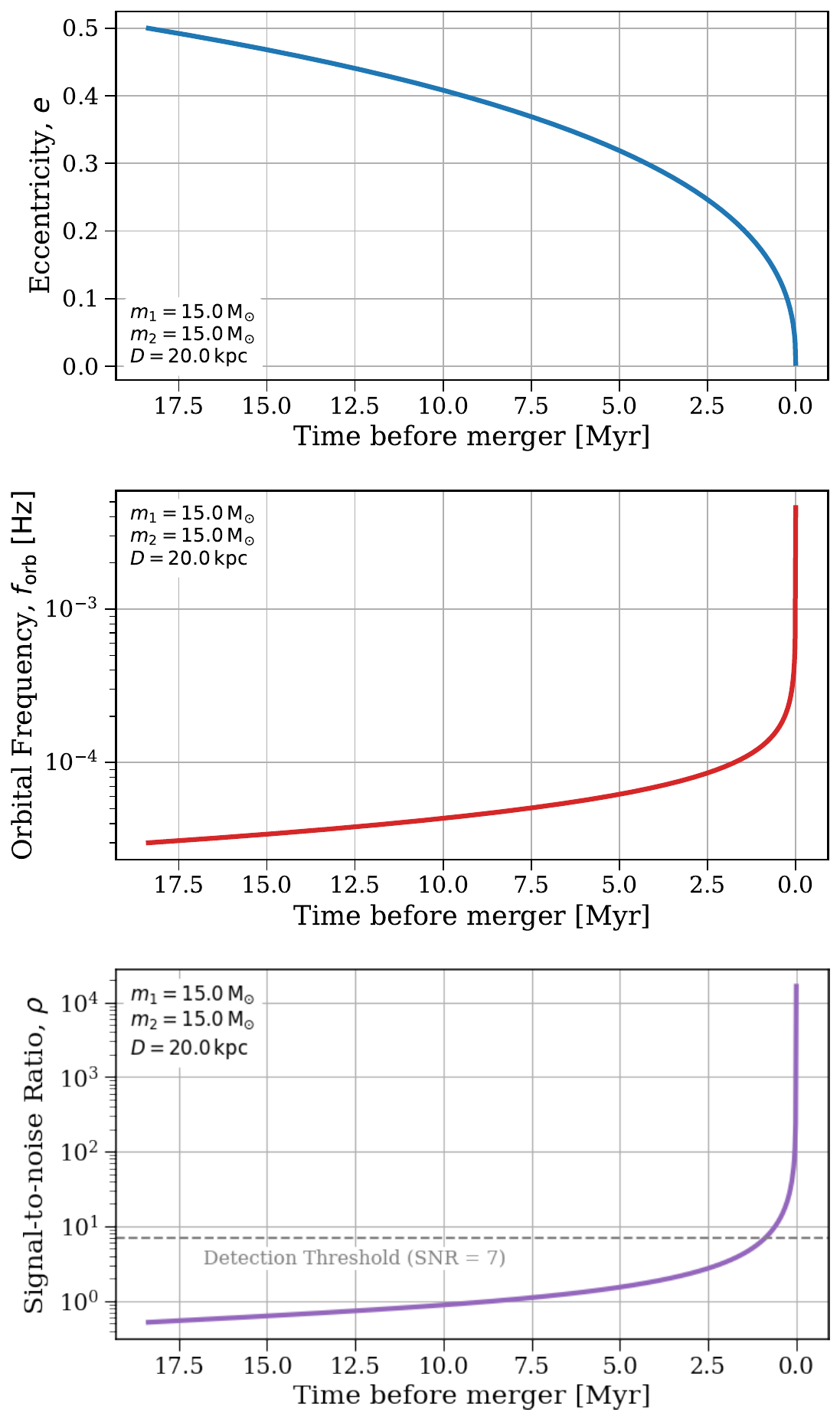}
    \caption{The evolution of a binary system's eccentricity (top), orbital frequency (middle) and SNR (bottom). Each panel is annotated with the constant parameters of the system. The SNR is calculated for a 4-year LISA mission. We annotate in  a line the bottom panel at $\mathrm{SNR} = 7$ to highlight the moment at which the source becomes detectable.}
    \label{fig:snr_over_time}
    \script{snr_over_time.py}
\end{figure}

We plot this evolution of the eccentricity and frequency in the top two panels of Figure~\ref{fig:snr_over_time} as a function of the time before the merger. We see that the binary circularises and increases its orbital frequency as it inspirals as we would expect.

To take this a step further, we can consider the binary at each timestep to be a separate source with the current eccentricity and frequency. It is then trivial to use \lw{} to calculate the SNR for each of these `sources' and thus attain the SNR evolution, which we plot in the last panel of Figure~\ref{fig:snr_over_time}.

We see that the SNR increases monotonically over time and sharply increases as the binary approaches its merger. Around $1 \unit{Myr}$ before the merger, the SNR reaches the detection threshold and thus could then be seen by a 4-year LISA mission.

Note that \lw{} could also be used in this way to find the SNR of any system in which the orbital evolution is known. Thus for a triple system or a binary experiencing gas drag (as long as the evolution of the eccentricity and orbital frequency is known) \lw{} is entirely capable of calculating the SNR evolution.

\subsection{LISA verification binaries\texorpdfstring{ - \tutorialLink{https://legwork.readthedocs.io/en/latest/demos/VerificationBinaries.html}}{}}
\citet{Kupfer+2018} presents the LISA verification binaries, a collection of known binary systems that have gravitational-waves that are strong enough to be detected by LISA. In \lw{}, we provide easy access to this data through the \href{https://legwork.readthedocs.io/en/latest/api/legwork.source.VerificationBinaries}{\color{\lwColour}{\texttt{VerificationBinaries}}} class, which is a subclass of \href{https://legwork.readthedocs.io/en/latest/api/legwork.source.Source}{\color{\lwColour}{\texttt{Source}}}. This means that the class works identically to \texttt{Source}, but it has the verification binary data (such as their masses and orbital frequencies) pre-loaded into the variables.

In addition to the base variables, this class also includes the designation of each binary and the SNR that \citet{Kupfer+2018} computed. We note that this SNR differs from the SNR that \lw{} gives for each binary. This is because \citet{Kupfer+2018} run a full, detailed LISA simulation for these binaries, whilst we follow the orbit-averaged approach (see Section~\ref{sec:amplitude_mod}), since the former would be intractable for any large number of sources.

As an example of how you could use this data, we use the \texttt{VerificationBinaries} class to plot the sources on the LISA sensitivity curve with the SNR calculated by \citet{Kupfer+2018} for a 4-year LISA mission and colour the binaries by their primary mass. We see that the verification binaries tend to be detected with frequencies between $1$-$10 \unit{mHz}$ and have masses estimated to be between $0.1$-$1 \unit{M_{\odot}}$.

\begin{figure}[htb]
    \centering
    \includegraphics[width=\columnwidth]{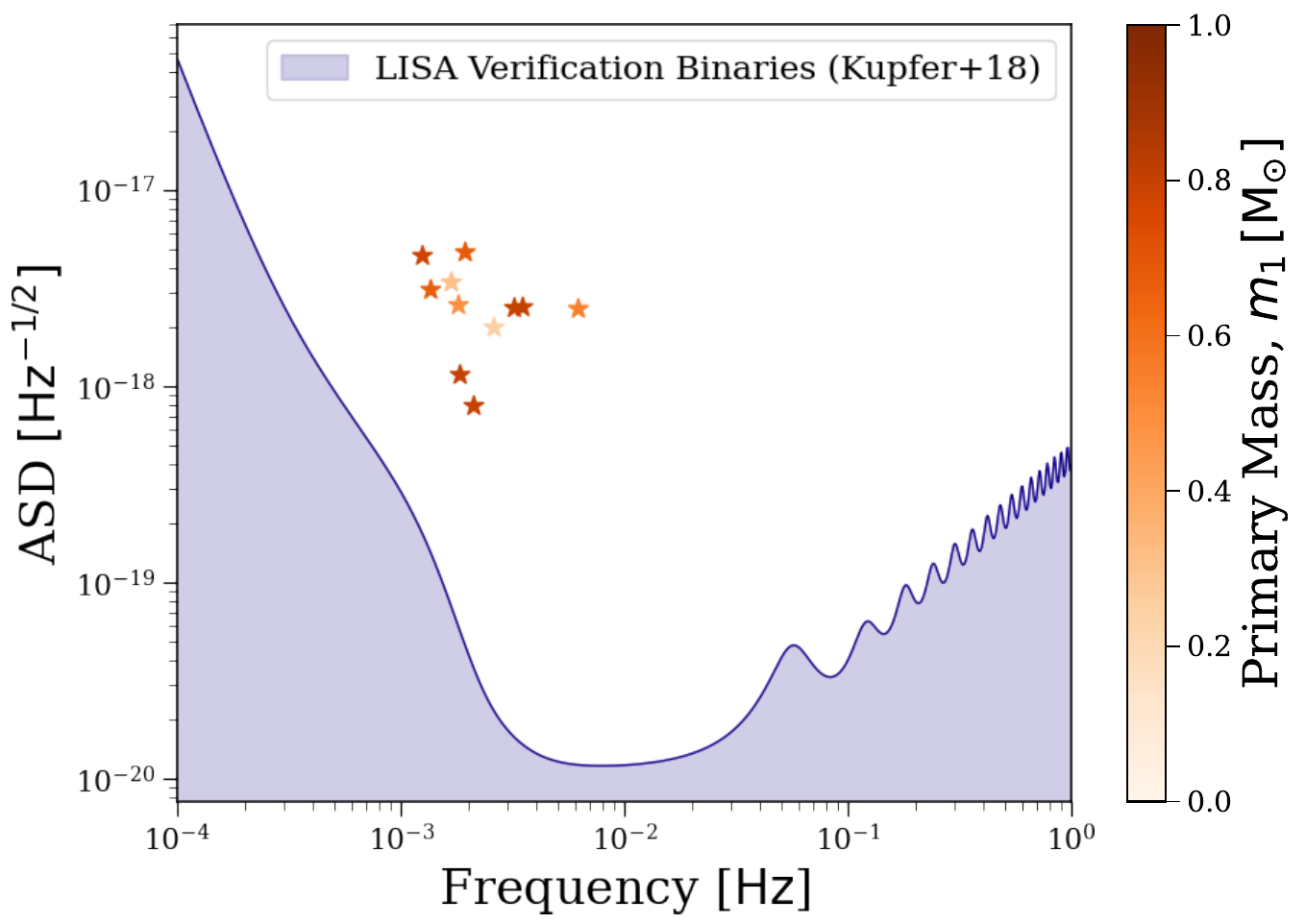}
    \caption{The LISA verification binaries from \citet{Kupfer+2018} plotted on the LISA sensitivity curve \citep{Robson+2019} for a 4-year LISA mission. Points are coloured by their primary mass.}
    \label{fig:verification_binaries_on_sc}
    \script{verification_binaries_on_sc.py}
\end{figure}

\section{Conclusion \& Summary}\label{sec:summary}
We have presented \lw{}, a package designed to aid in calculations for stellar-origin binary sources of mHz gravitational-wave observatories like LISA. We outlined the implementation of orbital evolution due to gravitational-wave emission, gravitational wave strain, SNR, and visualisation modules and provided a detailed derivation for each of the equations required for each module. Finally, we provided several use case examples for how \lw{} can be used to better understand the detectability of compact-object binaries.

\software{\lw{} is written in \texttt{Python}, available from \url{https://www.python.org}. We make use of the following Python packages: \texttt{matplotlib} \citep{Hunter+2007}, \texttt{NumPy} \citep{2020NumPy-Array},
\texttt{Astropy} \citep[\url{http://www.astropy.org}][]{AstropyCollaboration+2013,AstropyCollaboration+2018}, \texttt{Seaborn} \citep{Waskom+2021}, \texttt{SciPy} \citep{2020SciPy-NMeth}, \texttt{Numba} \citep{numba} and \texttt{Schwimmbad} \citep{Price-Whelan+2017}. This paper was compiled using \texttt{showyourwork} \citep{Luger+2021}.}

\begin{acknowledgements}
    We are grateful to Stas Babak, Floor Broekgaarden, Tom Callister, Will Farr, Yi-Ming Hu, Stephen Justham, Valeria Korol, Mike Lau, Tyson Littenberg, Ilya Mandel, Alberto Sesana, Lieke van Son, the CCA GW group, the BinCosmos group and the COMPAS group for stimulating discussions that influenced and motivated us to complete this project. We thank the BinCosmos group for testing an early version of the package and providing useful feedback. In particular, we thank Lieke van Son for her innovation in inventing the name \lw{}! TW thanks Floor Broekgaarden for first suggesting that he investigate LISA and the derivation of the SNR calculation. KB thanks Shane L. Larson and Kyle Kremer for \emph{several} conversations about the SNR derivation. We additionally thank the anonymous reviewer for their useful comments, which helped to improve the quality and clarity of this paper.
    
    This project was funded in part by the National Science Foundation under Grant No.\ (NSF grant number 2009131), the European Union’s Horizon 2020 research and innovation program from the European Research Council (ERC, Grant agreement No.\ 715063), and by the Netherlands Organization for Scientific Research (NWO) as part of the Vidi research program BinWaves with project number 639.042.728. We further acknowledge the Black Hole Initiative funded by a generous contribution of the John Templeton Foundation and the Gordon and Betty Moore Foundation. The Flatiron Institute is funded by the Simons Foundation.
\end{acknowledgements}

\bibliography{bib}
\bibliographystyle{aasjournal}

\end{document}